\documentclass{article}

\usepackage{arxiv}

\usepackage{multirow}
\usepackage[utf8]{inputenc} 
\usepackage[T1]{fontenc}    
\usepackage{hyperref}       
\usepackage{url}            
\usepackage{booktabs}       
\usepackage{amsfonts}       
\usepackage{nicefrac}       
\usepackage{microtype}      
\usepackage{lipsum}
\usepackage{graphicx}
\usepackage[normalem]{ulem}
\graphicspath{ {./images/} }
\usepackage{amsmath,amssymb}
\usepackage{xcolor}
\usepackage{graphicx,booktabs,url,array}
\usepackage{enumitem}
\usepackage{setspace}
\usepackage{wasysym}
\def\bSig\mathbf{\Sigma}

\renewcommand{\cite}{\citep}

\newcolumntype{L}[1]{>{\raggedright\arraybackslash}m{#1}}
\newcolumntype{C}[1]{>{\centering\arraybackslash}m{#1}}
\usepackage{amsmath, amsthm, amssymb, calrsfs, wasysym, verbatim, bbm, color, graphics, geometry, natbib}
\usepackage{amsthm}
\usepackage{bm}
\newtheorem{assumption}{Assumption}
\newtheorem{definition}{Definition}
\newtheorem{theorem}{Theorem}

\usepackage{CJKutf8}
\usepackage{soul,xcolor}
\setstcolor{red}
\renewcommand{\cite}{\citep}

\title{Evolving Longitudinal Patient Histories and Re-enrollment in Master Protocol Trials}

\author{
  Shiyu Wan$^{1}$, Yuhan Qian$^{1}$, Yanyao Yi$^{2}$, Nicole Mayer\mbox{-}Hamblett$^{3,4,1}$, Patrick J.~Heagerty$^{1}$, Ting Ye$^{1}$\\\vspace{2mm}\\
\small $^{1}$Department of Biostatistics, University of Washington\\
\small $^{2}$Global Statistical Sciences, Eli Lilly and Company\\
\small $^{3}$Seattle Children’s Research Institute \\
\small $^{4}$Department of Pediatrics, University of Washington
}
\renewcommand{\today}{}

\begin{document}
\maketitle
\begin{abstract}
A master protocol trial uses a single overarching protocol to test multiple therapies, often across several diseases or subtypes. Although such trials offer considerable flexibility and efficiency, their constrained and non-uniform treatment assignment raises two core challenges: precisely defining treatment effects and conducting robust, efficient inference. These challenges intensify when {participants} can re-enroll to receive additional eligible therapies over time. To address these issues,  we first define a clinically meaningful estimand with a clear population specification for master protocol trials  that allow re-enrollment across multiple episodes. Specifically, we define the episode-specific entire concurrently eligible (ECE) population, which preserves the integrity of randomized comparisons and remains invariant to randomization ratios and operational formats. We then introduce a per-episode added-{effect} estimand that aggregates episode-specific effects into an interpretable overall measure. For inference, we develop weighting and post-stratification estimators under the same minimal assumptions as conventional randomized trials, with model-assisted covariate adjustment to improve efficiency. We establish asymptotic distributions for all estimators and provide cluster-robust variance estimators that properly account for within-participant correlation induced by re-enrollment. We evaluate our methods through extensive simulations and apply our methods to SIMPLIFY, a master protocol trial comparing continuation versus discontinuation of two common cystic fibrosis therapies. All analyses are conducted using the \textsf{R} package \textsf{RobinCID}.
\end{abstract}

\textbf{Keywords:} Causal inference; Covariate adjustment; Estimands; Inverse probability weighting; Platform trials.

\section{Introduction}
\label{sec:intro}

Traditional approaches to generating high-quality medical evidence rely on a series of randomized controlled trials (RCTs), each typically evaluating one or two interventions within a single disease and a relatively homogeneous patient population \cite{berry2015platform}. However, as therapies become more targeted and patient populations more heterogeneous, this model has grown increasingly costly and operationally challenging.

In response, master protocol trials have emerged as overarching frameworks that facilitate parallel evaluation of one or more therapies across one or more diseases or patient subgroups under a single protocol, often using coordinated cohorts and shared controls \cite{woodcock2017master,fda2023masterprotocols}. Master protocol trials are commonly classified into three types: umbrella trials, which test multiple therapies in a single disease \cite{renfro2017statistical,rashdan2016battle}; basket trials, which test a single targeted therapy across multiple diseases or disease subtypes \cite{hobbs2022basket,Redig2015}; and platform trials, which continuously evaluate multiple therapies in an ongoing manner, allowing treatments to enter or leave the platform over time \cite{berry2015platform,saville2016efficiencies}. These basic types are not mutually exclusive; a single master protocol trial can incorporate features of more than one type. Although most early master protocols were developed in oncology \cite{woodcock2017master,park2019systematic}, the design has since been adopted across many therapeutic areas, including infectious diseases, neurodegenerative diseases, autoimmune disorders, and psychiatric conditions \cite{Wells2012MCTPlatform,Recovery_NEJM,GoldStefanM.2022Ptat,Foltynie2023MAMS_PD, honap2024basket}.

Unlike standard parallel-group trials, master protocol trials often share two key features. The first is \textit{constrained} and \textit{non-uniform} treatment assignment: constrained means some participants have zero probability of receiving certain treatments, while non-uniform refers to varying non-zero assignment probabilities based on participant characteristics, enrollment time, or treatment availability. The second is participant \textit{re-enrollment}: participants who complete one treatment period may later enroll in another. Re-enrollment arises naturally in master protocol trials because eligibility criteria frequently overlap across arms and new arms may open over time while others close. As a result, a single patient can contribute multiple correlated observations, complicating both the definition and estimation of treatment effects. 

Two recent trials illustrate these features.  In the Lung-MAP (S1400) master protocol, {individuals with squamous non-small-cell lung cancer} were first allocated to one of six biomarker-matched substudies or, if no match was identified, to one of three non-match substudies. Upon disease progression, {participants} could re-enroll in a different substudy \cite{Redman2020LungMAP_S1400}.  In the SIMPLIFY trial, {individuals with cystic fibrosis} were assigned to either the hypertonic saline (HS) or dornase alfa (DA) substudy based on their baseline use of these therapies; after completing their initial substudy, eligible participants could subsequently enroll in the other substudy \cite{MayerHamblett2021SIMPLIFY_Design}. These design elements, including constrained and non-uniform assignment coupled with re-enrollment, distinguish master protocols from standard trials. They necessitate estimands that differ from those in conventional parallel-group designs and require tailored identification strategies and analytical methods. 

A growing body of literature addresses statistical methods for master protocol trials. Umbrella trials have received relatively little methodological attention, with separate analyses for each substudy remaining the standard approach \citep{Ouma2022UmbrellaTrials}. However, \citet{qian2025} recently demonstrated that this conventional practice can yield estimands that unexpectedly depend on randomization ratios, making them arbitrary and difficult to interpret. For basket trials, methodological developments have focused on adaptive design \cite{cunanan2017efficient} and Bayesian approaches that borrow information across subgroups \citep{chu2018bayesian, jin2020bayesian}. Statistical methods for platform trials have addressed both design aspects—including adaptive randomization \citep{yuan2016midas,kaizer2018multi,ventz2018adding} and optimization of allocation ratios \citep{bofill2024optimal}—and analytical considerations, particularly the use of nonconcurrent controls \citep{saville_bayesian_2022, bofill2023use,roig2022model, jiang2023nonconcurrent, wang_bayesian_2023, Liu2023BorrowingConcurrent, guo2024treatment, platform_nonconcurrent}.

More recently, researchers have begun to carefully consider how to define clinically meaningful estimands in master protocol trials. \citet{platform_nonconcurrent} focused on platform trials and introduced the concurrent average treatment effect, defined as the average treatment effect among concurrent units only. \citet{qian2025} considered a more general master protocol setting and proposed the entire concurrently eligible (ECE) population along with an associated estimand for treatment comparisons. However, no work has yet addressed estimands or robust estimation methods for master protocols with re-enrollment.

Our contributions are threefold. First, we develop a statistical framework for master protocol trials with re-enrollment, establishing clear assumptions that characterize the randomization mechanism under re-enrollment. Second, we define a clinically meaningful estimand tailored to the design features of constrained and non-uniform assignment coupled with re-enrollment. Third, we construct consistent estimators without relying on strong or untestable assumptions, and derive robust variance estimators that account for within-participant correlation induced by re-enrollment.

The remainder of the article is organized as follows. Section \ref{sec::setup} introduces the motivating trial, establishes notation, and defines the target estimands. Section \ref{sec::estimation} develops robust estimation methods and derives asymptotic distributions for the proposed estimators. Section \ref{sec::simulation} evaluates finite-sample performance through simulation studies. Section \ref{sec::real} applies the proposed methods to the SIMPLIFY trial. Section \ref{sec::discuss} concludes with a discussion. 

\section{Motivating Trial, Setup, and Estimand}\label{sec::setup}

\subsection{The SIMPLIFY trial} \label{sec::setup_simplify}

We begin by presenting the master protocol trial that motivates this work: the SIMPLIFY trial \cite{MayerHamblett2023SIMPLIFY}. Despite its apparently simple design, this trial embodies the core challenges characteristic of master protocol trials: constrained and non-uniform treatment assignment coupled with participant re-enrollment.

The SIMPLIFY trial evaluated whether people with cystic fibrosis taking the novel triple combination therapy elexacaftor/tezacaftor/ivacaftor (ETI) can {subsequently} safely discontinue pre-existing airway-clearance drugs, specifically hypertonic saline (HS) or dornase alfa (DA). The trial comprises two concurrent, multicenter, randomized controlled substudies, each assessing whether discontinuing one of these therapies is non-inferior to continuation. The primary outcome for this trial was the 6-week mean absolute change in percent-predicted forced expiratory volume in one second ($\text{ppFEV}_1$), with discontinuation hypothesized to be non-inferior to continuation using a -3\% non-inferiority margin.

\begin{figure}[htbp] 
    \centering
    \includegraphics[width=\columnwidth]{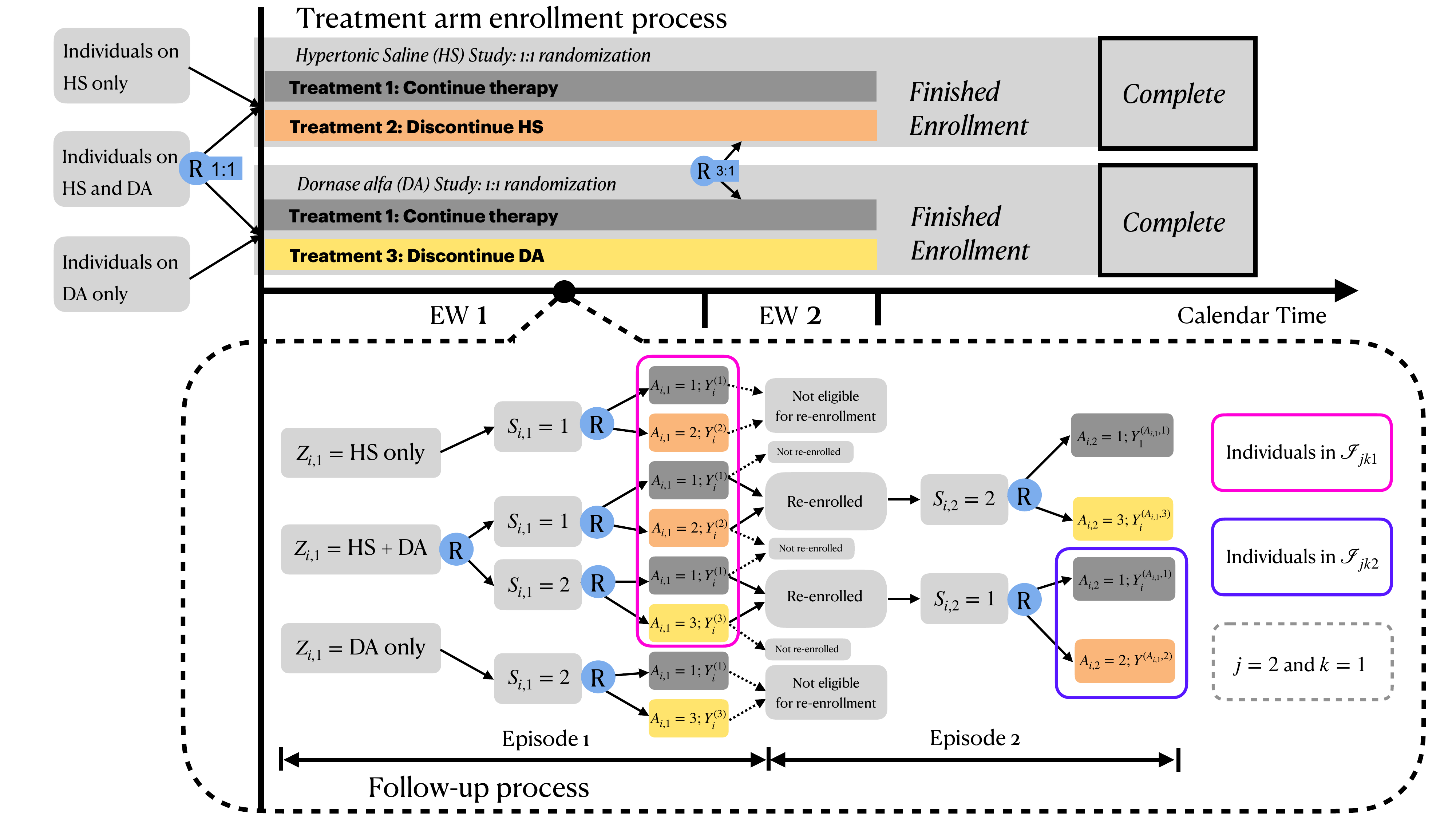}
    \caption{Illustration of SIMPLIFY treatment-arm enrollment and follow-up. Participants are first randomized to a substudy based on their baseline medication group, then randomized to the corresponding treatment arm. Randomization probabilities change in Enrollment Window 2 (EW2). After completing the first substudy, eligible participants may re-enroll in the alternate substudy.}
    \label{fig:SIMPLIFY_illustration}
\end{figure}

Figure~\ref{fig:SIMPLIFY_illustration} shows the SIMPLIFY study design: the top panel illustrates the enrollment process, and the bottom panel depicts follow-up for one enrolled individual. All enrolled individuals had received ETI for at least 90 days and fell into one of three baseline medication groups: HS only, DA only, or both HS and DA. Individuals using a single therapy were deterministically assigned to the corresponding substudy, while those using both therapies were randomized to either the HS or DA substudy. This randomization used a 1:1 ratio during enrollment window 1 (before January 14, 2022) and a 3:1 ratio during window 2 (on or after January 14, 2022) to allocate more participants to the HS substudy due to faster accrual in the DA substudy. Within each substudy, participants were randomized 1:1 to continue or discontinue the corresponding therapy. After completing their first substudy, participants who were using both therapies at baseline could re-enroll in the other substudy if still eligible and willing, since they had a second therapy that could be evaluated. In contrast, participants using only one therapy at baseline had no additional therapy to evaluate after their initial substudy. Approximately 58\% of participants on both HS and DA re-enrolled for a second episode, with similar re-enrollment rates regardless of initial substudy assignment. Table~\ref{tab:example} summarizes the assignment probabilities.

\begin{table}[ht]
\centering
\caption{Assignment probabilities used in the SIMPLIFY study and the simulation study}
\label{tab:example} 
\vspace{0.2cm}

\resizebox{\columnwidth}{!}{
\begin{tabular}{c|ccc|c|c|cc|ccc}
\toprule
\multicolumn{1}{c|}{\textbf{Episode}} &
\multicolumn{3}{c|}{\textbf{Z}} &
\multicolumn{1}{c|}{\textbf{T}} &
\multicolumn{1}{c|}{\textbf{History}} &
\multicolumn{2}{c|}{\textbf{Substudy}} &
\multicolumn{3}{c}{\textbf{Treatment}} \\
\cmidrule(lr){1-1}
\cmidrule(lr){2-4} \cmidrule(lr){5-5} \cmidrule(lr){6-6} \cmidrule(lr){7-8}  \cmidrule(lr){9-11}
$t$& $\text{HS}_{i}$ & $\text{DA}_{i}$ & $\mathrm{EW}_i$  & $T_i$ & $\pmb{\bar{S}}_{i,t-1}$ &
${S}_{i,t} = \text{HS}$ & ${S}_{i,t} = \text{DA}$ &
$\pi_1^t(Z_{i,t})$ & $\pi_2^t(Z_{i,t})$ & $\pi_3^t(Z_{i,t})$ \\
\midrule
1 & 1 & 0 & any & 1 & $\emptyset$ & 1.00 & 0.00 & 0.50 & 0.50 & 0.00 \\
1 & 0 & 1 & any & 1 & $\emptyset$ & 0.00 & 1.00 & 0.50 & 0.00 & 0.50 \\
1 & 1 & 1 & 1   & any & $\emptyset$ & 0.50 & 0.50 & 0.50 & 0.25  & 0.25 \\
1 & 1 & 1 & 2   & any & $\emptyset$ & 0.75 & 0.25 & 0.50 & 0.375 & 0.125 \\
2 & 1 & 1 & any & 2 & $\{\text{DA}\}$ & 1.00 & 0.00 & 0.50 & 0.50 & 0.00 \\
2 & 1 & 1 & any & 2 & $\{\text{HS}\}$ & 0.00 & 1.00 & 0.50 & 0.00 & 0.50 \\
\bottomrule
\end{tabular}
}

\vspace{0.15cm}
{\footnotesize
\textit{Note:} In the simulation,
$X_{i,(cat)}=0$ corresponds to $(\mathrm{HS}_i,\mathrm{DA}_i)=(1,0)$,
$X_{i,(cat)}=1$ corresponds to $(\mathrm{HS}_i,\mathrm{DA}_i)=(0,1)$,
and $X_{i,(cat)}=2$ corresponds to $(\mathrm{HS}_i,\mathrm{DA}_i)=(1,1)$.
Moreover, $X_{i,\mathrm{EW}}$ corresponds to $\mathrm{EW}_i$.
Thus, the $Z$ columns in this table correspond to the simulation covariates
$Z_{i,t}=\bigl(X_{i,(cat)},\,X_{i,\mathrm{EW}}\bigr)$ via this mapping.
}

\end{table}

\subsection{Setup}
Consider a master protocol trial with $n$ individuals, $J$ treatment arms, $S$ substudies, and a maximum of $T$ episodes per individual, where an \textit{episode} is a single instance of an individual being enrolled in a treatment arm. For individual $i \in \{1,\cdots,n\}$ and episode $t \in \{1,\cdots,T\}$, we define the following quantities. Let $\pmb{W}_{i,t}$ denote the vector of baseline covariates measured before treatment assignment at episode $t$, including randomization factors such as enrollment window and other baseline covariates. Let $S_{i,t}$ denote the substudy assignment indicator that equals $s$ if individual $i$ is assigned to substudy $s$ at episode $t$. Let $A_{i,t}$ denote the treatment assignment indicator that equals $j$ if individual $i$ is assigned to arm $j$ at episode $t$. Let $Y_{i,t}$ denote the observed outcome.  If the master protocol does not include substudies, $S_{i,t}$ can be omitted.

For $t\geq 1$, let ${\pmb{\bar{W}}}_{i,t} = ({\pmb{W}}_{i,1},\cdots,{\pmb{W}}_{i,t})$, ${\pmb{\bar{S}}}_{i,t} = (S_{i,1},\cdots,S_{i,t})$, ${\pmb{\bar{A}}}_{i,t} = (A_{i,1},\cdots,A_{i,t})$, and ${\pmb{\bar{Y}}}_{i,t} = (Y_{i,1},\cdots,Y_{i,t})$ denote the histories of baseline covariates, substudy assignments, treatments assignments, and observed outcomes, respectively, for individual $i$ through episode $t$.  By convention, when $t=0$, we set $\pmb{\bar{W}}_{i,0} = \pmb{\bar{S}}_{i,0} = \pmb{\bar{A}}_{i,0} = {\pmb{\bar{Y}}}_{i,0} = \emptyset$. 

Let $T_i$ denote the total number of treatment episodes attended by individual $i$. For each individual $i$, we define the vectors of baseline covariates, substudy assignments, treatment assignments, and observed outcomes across all attended episodes as $\pmb{W}_i := \pmb{\bar{W}}_{i,T_i}= (\pmb{W}_{i,1},\cdots,\pmb{W}_{i,T_i})$, $\pmb{S}_i := \pmb{\bar{S}}_{i,T_i} = (S_{i,1},\cdots,S_{i,T_i})$, $\pmb{A}_i := \pmb{\bar{A}}_{i,T_i} = (A_{i,1},\cdots,A_{i,T_i})$, and $\pmb{Y}_i := {\pmb{\bar{Y}}}_{i,T_i} =(Y_{i,1},\cdots,Y_{i,T_i})$. For individuals with $T_i \geq t$, let $\mathcal{F}_{i,t-1} = \{\pmb{\bar{W}}_{i,t},\pmb{\bar{S}}_{i,t-1},\pmb{\bar{A}}_{i,t-1},\pmb{\bar{Y}}_{i,t-1}\}$ denote the observed history prior to treatment assignment at episode $t$.

Let $\mathcal{A}_{i,t}$ denote the set of all possible treatment histories for individual $i$ through episode $t$. We
write $a_t$ for a realization of $A_{i,t}$ and $\bar{a}_t = (a_1,\cdots,a_t) \in \mathcal{A}_{i,t}$ for a realization of $\pmb{\bar{A}}_{i,t}$. For any treatment history $\bar{a}_t \in \mathcal{A}_{i,t}$, let $Y_{i,t}^{(\bar{a}_t)}$ denote the potential outcome for individual $i$ at episode $t$ under treatment history  $\bar{a}_t$. The collection $\{Y_{i,t}^{(\bar{a}_t)}: \bar{a}_t \in \mathcal{A}_{i,t}\}$ represents all potential outcomes at episode $t$. Under the consistency assumption, the observed outcome satisfies $Y_{i,t} = Y_{i,t}^{(\pmb{\bar{A}}_{i,t})}$. We assume the individual-level data $(\pmb{W}_i,\pmb{A}_i,\pmb{S}_i,T_i, \{Y_{i,t}^{(\bar{a}_t)}: \bar{a}_t \in \mathcal{A}_{i,t}\}_{t=1}^{T_i})$ are mutually independent across individuals.

\begin{assumption}[Sequential Treatment Randomization] \label{assump: rand}
    For each individual $i$ and episode $t \in \{1,\dots, T_i\}$, let $Z_{i,t} \subseteq \mathcal{F}_{i,t-1}$ denote the subset of observed history used to determine eligibility and treatment assignment at episode $t$. We assume: \\
    (i) For all ${a}_{t}$ such that $(\bar{\pmb{A}}_{i,t-1},a_t)\in \mathcal{A}_{i,t}$, we have $A_{i,t} \perp  {Y}_{i,t}^{(\bar{\pmb{A}}_{i,t-1},a_t)} \mid \mathcal{F}_{i,t-1}, T_i\geq t$.\\
    (ii) The treatment assignment probability depends on the observed history only through $Z_{i,t}$: $\mathbb{P}(A_{i,t} = j|\mathcal{F}_{i,t-1},T_i\geq t)  = \mathbb{P}(A_{i,t} = j|Z_{i,t},T_i\geq t) =\pi_{j}^{t}(Z_{i,t})$, where $0 \leq  \pi_j^t (Z_{i,t}) \leq 1 $ are known for all $ j = 1,\cdots,J$, and $\sum_{j=1}^J \pi_j^t (Z_{i,t}) = 1$ for all $ t\leq T_i$.
\end{assumption} 

Assumption \ref{assump: rand} formalizes the sequential randomization structure inherent in master protocol trials where individuals may enroll in multiple treatment episodes. Assumption \ref{assump: rand}(i) states that, conditional on the observed history $\mathcal{F}_{i,t-1}$ and attending episode $t$ ($T_i \geq t$), the treatment assignment $A_{i,t}$ is independent of potential outcomes under any feasible treatment history. Assumption \ref{assump: rand}(ii) requires that treatment assignment probabilities at each episode depend on observed history only through a known subset $Z_{i,t}$ and that these probabilities are known. Both conditions are satisfied by design in master protocol trials with pre-specified randomization schemes, where assignment probabilities are determined by the trial protocol and may depend on factors such as enrollment window, prior substudy and treatment history, and individual characteristics. {For example, in the SIMPLIFY trial, randomization at episode 1 depends on baseline use of HS and DA and on enrollment window. Only individuals using both HS and DA at their first entry into the master protocol are eligible for re-enrollment. For these individuals, randomization at episode 2 depends on their prior substudy assignment. All randomization probabilities are specified by the trial protocol.} A directed acyclic graph (DAG) representing Assumption~\ref{assump: rand} is shown in Figure~\ref{fig:SIMPLIFY_DAG}.


\begin{figure}[htbp] 
    \centering
    \includegraphics[width=\columnwidth]{DAG_draft_V2.pdf}
    \caption{Illustration of the SIMPLIFY DAG. The red path highlights the collider path opened by conditioning on re-enrollment, which is blocked by further conditioning on $Z_{i,2}$.}
    \label{fig:SIMPLIFY_DAG}
\end{figure}

Importantly, Assumption \ref{assump: rand} imposes no restrictions on the re-enrollment mechanism.  Even if unmeasured factors influence re-enrollment decisions, the randomization of treatment assignment among re-enrolled individuals ensures identification of treatment contrasts. This is analogous to traditional trials, where no assumptions are required about participant selection into the study, yet treatment effects remain identifiable through randomization. We show in Section \ref{sec::estimation} that Assumption \ref{assump: rand} alone suffices to identify per-episode treatment effects.

\subsection{Estimand}\label{sec::estimand}

How to precisely define meaningful estimands in master protocol trials without re-enrollment has been extensively discussed by \citet{qian2025}. They proposed the entire concurrently eligible (ECE) population as the basis for treatment comparisons, arguing that this definition preserves the integrity of randomized comparisons, is invariant to randomization ratios and platform trial formats, and aligns with the intention-to-treat principle \citep{ICH_E9_1998_FDA}. The ECE population is the same as the entire trial population in traditional trials but generalizes naturally to handle the complications of master protocol trials. In contrast, populations defined by actual treatment received or substudy allocation depend on the randomization ratio and can therefore be arbitrary. In this section, we discuss estimands in the presence of re-enrollment. 

First, we define the episode-specific ECE population for comparing two treatment {regimens} $j$ and $k$ (one of which is typically a control). In this paper, ``population'' specifically refers to the population of individuals used to define the estimand.

\begin{definition}[Episode-$t$ ECE Population] 
For treatment {regimens} $j$ and $k$, the episode-$t$ ECE population is a population of all individuals who could potentially be enrolled for episode $t$ during a time period when both treatment regimens are available and meet the eligibility criteria for both treatment regimens at this episode, and therefore could be randomized to either treatment at episode $t$. Formally, this is the population of all individuals with $ \pi_j^{t}(Z_{i,t}) > 0, \pi_k^{t}(Z_{i,t}) > 0$, and $T_i \geq t$. 
\end{definition}

The episode-$t$ ECE population is a population of all individuals who could \emph{potentially} be assigned to treatment {regimens} $j$ and $k$ at episode $t$, rather than the population represented by those \emph{actually} assigned to these treatments (i.e., the population represented by $A_{i,t}\in \{j, k\}$). For example, in the SIMPLIFY trial, the episode-1 ECE population for comparing treatments 1 and 2 (i.e., evaluating the effect of discontinuing HS versus continuing) includes all individuals on HS only or on both HS and DA at baseline. The episode-2 ECE population for comparing treatments 1 and 2 includes all individuals on both HS and DA at baseline, entered the DA study at episode 1, and were re-enrolled at episode 2, as shown in Figure~\ref{fig:SIMPLIFY_illustration}.

We then propose the following estimand to define the treatment effect for treatment {regimens} $j$ and $k$ within the episode-$t$ ECE population: 
\begin{align*}
    \pmb{\vartheta}_{jkt} = 
    \left (
        \begin{matrix}
            \theta_{jkt} \\
            \theta_{kjt}
        \end{matrix}
    \right ) = \left (
        \begin{matrix}
            \mathbb{E}(Y_{i,t}^{(\bar{\pmb{A}}_{i,t-1},j)}\mid i \in \mathcal{I}_{jkt}) \\
            \mathbb{E}(Y_{i,t}^{(\bar{\pmb{A}}_{i,t-1},k)}\mid i \in \mathcal{I}_{jkt})
        \end{matrix}
    \right ),
\end{align*}
where  $\mathcal{I}_{jkt} = \{i: \pi_j^{t}(Z_{i,t}) > 0, \pi_k^{t}(Z_{i,t}) > 0, \text{ and }T_i \geq t\}$. Here $\theta_{jkt}$ and $\theta_{kjt}$ depend on $j,\;k$, and $t$ because the ECE population depends on them, and the expectation averages over the distribution of pre-episode-$t$ treatment histories. A contrast of $\theta_{jkt}$ and $\theta_{kjt}$ is an unconditional treatment effect for the episode-$t$ ECE population, representing the additional effect of the treatment assigned in the current episode beyond any {effect that may have} accumulated from prior episodes. This parallels the estimand from a traditional parallel-group trial: it evaluates assignment to treatment $j$ versus $k$ over the entire concurrently eligible population at episode $t$,  averaging over individual characteristics and past treatment histories that may have been shaped by earlier treatment assignments.

Next, we consider how to aggregate episode-specific treatment effects into an overall measure. As noted by \citet{Kahan2022Rerandomisation}, estimands in multi-episode settings depend on two main considerations: how to handle prior treatment history and how to aggregate episode-specific effects. For treatment history, one can {evaluate whether there is} (i) \emph{added-{effect}}, which marginalizes over prior histories, or (ii) \emph{policy-{effect}}, which contrasts full treatment policies such as ``always $j$" versus ``always $k$." For aggregation, one can average across episodes (\emph{per-episode}) or across participants (\emph{per-participant}). Combining these two considerations yields four estimands.

The \emph{per-episode added-{effect}} estimand averages episode-specific treatment effects across all episodes, weighting each by the probability of belonging to that episode's ECE population. (i.e., each episode contributes equally) and marginalizing over prior treatment histories within each episode. The \emph{per-participant added-{effect}} estimand first computes each {participant}'s average treatment effect across their episodes, then averages these participant-level means equally across {participants}, so that each {participant} contributes equally regardless of how many episodes they have. The two \emph{policy-{effect}} estimands—per-episode and per-participant—contrast full treatment policies (e.g., ``always $j$" versus ``always $k$") using the corresponding weighting schemes. In re-enrollment trials, the per-episode added-{effect} estimand can generally be estimated under minimal assumptions, requiring only that treatment assignment be randomized within each episode conditional on observed variables. 
The per-participant added-{effect} estimand involves weights based on the number of episodes each individual attends, which is a random quantity that complicates interpretation. The two policy-{effect} estimands face additional challenges: identifying the effect of a full treatment policy requires assumptions about re-enrollment decisions, yet {participants}' decisions to re-enroll are likely influenced by unmeasured factors such as prior outcomes and experiences, making it difficult to identify full-policy effects without strong, untestable assumptions.

Therefore, to ensure interpretability and identifiability under minimal assumptions, we consider \emph{per-episode added-{effect}} as the primary overall estimand, defined as:
\begin{align}
    \pmb{\vartheta}_{jk} = 
    \left (
        \begin{matrix}
            \theta_{jk} \\
            \theta_{kj}
        \end{matrix}
    \right ) = \left (
        \begin{matrix}
             \frac{\sum_{t=1}^T\mathbb{P}(I_{jkt}(i) = 1)\theta_{jkt}}{\sum_{t=1}^T\mathbb{P}(I_{jkt}(i) = 1)} \\
             \frac{\sum_{t=1}^T \mathbb{P}(I_{jkt}(i) = 1)\theta_{kjt}}{\sum_{t=1}^T\mathbb{P}(I_{jkt}(i) = 1)}
        \end{matrix}
    \right ),\label{eq: estimand}
\end{align}
where $I_{jkt}(i)= I\{\pi_{j}^t(Z_{i,t}) >0,\pi_{k}^t(Z_{i,t}) > 0,T_i \geq t\}$ indicates membership in the episode-$t$ ECE population. Each component of $  \pmb{\vartheta}_{jk}$ is a weighted average of episode-specific treatment effects, with weights proportional to the probability of belonging to each episode's ECE population. Equivalently, this estimand represents the average treatment effect for a randomly selected person-episode from the pooled ECE population across all episodes, so that each person-episode contributes equally to the overall estimand. A linear contrast such as $\theta_{jk} -\theta_{kj}$ is therefore a weighted average of the episode-specific unconditional treatment effects.
In the SIMPLIFY trial, $\theta_{21} -\theta_{12}$ is a weighted average of two “discontinue HS versus continue” effects: one evaluated in the episode-1 ECE population (individuals on HS alone or HS+DA at baseline), and one evaluated in the episode-2 ECE population (individuals on HS+DA at baseline who re-enrolled in the HS study after episode 1). The weights reflect the probability of belonging to each ECE population.

\section{Estimation and Inference}\label{sec::estimation}

In this section, we develop consistent and asymptotically normal estimators for the proposed estimand in \eqref{eq: estimand} under minimal assumptions, along with robust variance estimators that account for within-participant correlation induced by re-enrollment.

\subsection{Estimation}

In master protocol trials, treatment assignment depends on $Z_{i,t}$, so naively averaging $Y_{i,t}$ among individuals from the episode-$t$ ECE population with $A_{i,t} = j$ is generally biased for $\theta_{jkt}$. We therefore use weighting or post-stratification to adjust for confounding by $Z_{i,t}$.

\subsubsection{Weighting}

For any episode $t$, the inverse probability weighting (IPW) estimator for  $\theta_{jkt}$ is:
\begin{align*}
    \hat{\theta}_{jkt}^{\text{ipw}} = \frac{1}{|\mathcal{I}_{jkt}|} \sum_{i \in \mathcal{I}_{jkt}} \frac{I(A_{i,t} = j)Y_{i,t}}{\pi_{j}^t(Z_{i,t})}
\end{align*}
where  $\mathcal{I}_{jkt} = \{i: \pi_j^{t}(Z_{i,t}) > 0, \pi_k^{t}(Z_{i,t}) > 0, \text{ and }T_i \geq t\}$ is defined as before and $|\mathcal{I}_{jkt}|$ denotes its cardinality. To estimate $\theta_{jk}$, we average the episode-specific IPW estimators $\{\hat{\theta}^{\text{ipw}}_{jkt}\}_{t=1}^T$, weighting each by $|\mathcal{I}_{jkt}|$:
\begin{align*}
    \hat{\theta}_{jk}^{\text{ipw}}& = \frac{1}{n_{jk}}  \sum_{t = 1}^T  |\mathcal{I}_{jkt}| \ \hat{\theta}_{jkt}^{\text{ipw}}  = \frac{1}{n_{jk}}  \sum_{t = 1}^T  \sum_{i \in \mathcal{I}_{jkt}}   \frac{I(A_{i,t} = j)Y_{i,t}}{\pi_{j}^t(Z_{i,t})}, 
\end{align*}
where $n_{jk}= \sum_{t=1}^T |\mathcal{I}_{jkt}|$ is the total number of person-episodes across all episodes. The rightmost expression shows that $\hat{\theta}_{jk}^{\text{ipw}}$ can be viewed as a weighted average of outcomes across all person-episodes, with each person-episode contributing equally after applying IPW. 

The IPW estimator can be improved by the stabilized IPW (SIPW) \cite{Hajek1971-comment-essay}, which normalizes the weights to sum to one:
\begin{align*}
    \hat{\theta}_{jk}^{\text{sipw}}& = \left\{   \sum_{t = 1}^T  \sum_{i \in \mathcal{I}_{jkt}}   \frac{I(A_{i,t} = j)}{\pi_{j}^t(Z_{i,t})} \right\}^{-1}  \sum_{t = 1}^T  \sum_{i \in \mathcal{I}_{jkt}}   \frac{I(A_{i,t} = j)Y_{i,t}}{\pi_{j}^t(Z_{i,t})}.
\end{align*}
Both the IPW and SIPW estimators are consistent and asymptotically normal (Theorem \ref{theo: CAN}).

To improve the efficiency of the IPW and SIPW estimators, we can adjust for $Z_{i,t}$ and other covariates using a model-assisted approach \cite{Ye02102023,robincar}. This approach incorporates covariates through a working outcome model and yields an asymptotically unbiased estimator even if the working model is misspecified. Let $\pmb{X}_{i,t} \subset \pmb{W}_{i,t}$ denote the observed vector of baseline covariates for episode $t$ used for adjustment, which may overlap with $Z_{i,t}$. For each $t=1,\dots, T$,
we estimate $\mathbb{E}(Y_{i,t}|\pmb{X}_{i,t},I_{jkt}(i)=1,A_{i,t} = j)$ by $\hat\mu_{jkt}(\pmb{X}_{i,t})$ using a working model that may be misspecified. Since the working model need not be correctly specified, one can either fit separate models using data from each $\mathcal{I}_{jkt}$ with $A_{i,t}= j$, or fit a single model using data pooled across all episodes to reduce complexity. We then adjust for covariates using augmented inverse probability weighting (AIPW) \citep{robins1994estimation}:
\begin{align*}
    \hat{\theta}_{jk}^{\text{aipw}}= \frac{1}{n_{jk}}  \sum_{t = 1}^T  \sum_{i \in \mathcal{I}_{jkt}} \frac{I(A_{i,t} = j)\{Y_{i,t}-\hat{\mu}_{jkt}(\pmb{X}_{i,t})\}}{\pi_{j}^t(Z_{i,t})}+\hat{\mu}_{jkt}(\pmb{X}_{i,t}).
\end{align*}

\subsubsection{Stratification}

When $\pi_{j}^t(Z_{i,t})$ and $\pi_{k}^t(Z_{i,t})$ take discrete values, an alternative to IPW is post-stratification (PS), which stratifies individuals based on the values of $\pi_{j}^t(Z_{i,t})$  and $\pi_{k}^t(Z_{i,t})$. Specifically, we divide all concurrently eligible individuals at episode $t$ (i.e., those in $\mathcal{I}_{jkt}$) into a finite number of strata $\mathcal{I}_{jkt}^{(1)},\cdots,\mathcal{I}_{jkt}^{(H_{jkt})}$, where $H_{jkt}$ is the number of strata at episode $t$, such that within each stratum $h \in\{1,\cdots,H_{jkt}\}$, the values of $\pi_{j}^t(Z_{i,t})$ and $\pi_{k}^t(Z_{i,t})$ are constant for all $i\in \mathcal{I}_{jkt}^{(h)}$. The PS estimator of $\theta_{jkt}$ is:
\begin{align*}
    \hat{\theta}^{\text{ps}}_{jkt} = \frac{1}{|\mathcal{I}_{jkt}|}&\sum_{h=1}^{H_{jkt}}\frac{n(\mathcal{I}_{jkt}^{(h)})}{n_j(\mathcal{I}_{jkt}^{(h)})} \sum_{i\in \mathcal{I}_{jkt}^{(h)} }I(A_{i,t} = j) Y_{i,t},
\end{align*}
where $n(\mathcal{I}_{jkt}^{(h)})$ is the number of individuals in $\mathcal{I}_{jkt}^{(h)}$ and $n_j(\mathcal{I}_{jkt}^{(h)})$ is the number of individuals in $\mathcal{I}_{jkt}^{(h)}$ with $A_i = j$. The PS estimator of $\theta_{jk}$ is the weighted average of the episode-specific PS estimators $\{\hat{\theta}_{jkt}^{\text{ps}}\}_{t=1}^T$:
\begin{align*}
    \hat{\theta}^{\text{ps}}_{jk} =  \frac{1}{n_{jk}}\sum_{t=1}^T\sum_{h=1}^{H_{jkt}}\frac{n(\mathcal{I}_{jkt}^{(h)})}{n_j(\mathcal{I}_{jkt}^{(h)})} \sum_{i\in \mathcal{I}_{jkt}^{(h)} }I(A_{i,t} = j) Y_{i,t}.
\end{align*}

To adjust for additional covariates beyond the stratification variable, we apply AIPW within each stratum, yielding the adjusted post-stratification (APS) estimator: 
\begin{align*}
    \hat{\theta}^{\text{aps}}_{jk} &=  \frac{1}{n_{jk}}\sum_{t=1}^T\sum_{h=1}^{H_{jkt}}\frac{n(\mathcal{I}_{jkt}^{(h)})}{n_j(\mathcal{I}_{jkt}^{(h)})} \sum_{i\in \mathcal{I}_{jkt}^{(h)} }I(A_{i,t} = j) \{Y_{i,t}-\hat{\mu}_{jkt}(\pmb{X}_{i,t})\} + \frac{1}{n_{jk}}\sum_{t=1}^T\sum_{i\in \mathcal{I}_{jkt}} \hat{\mu}_{jkt}(\pmb{X}_{i,t}).
\end{align*}

\subsection{Asymptotic Theory}

To develop the asymptotic theory for estimators of $\bm\vartheta_{jk}$ with covariate adjustment, we require the following standard condition that the estimated working model $\hat\mu_{jkt}$ converges to a well-defined limit $\mu_{jkt}$, 
which need not be correctly specified.

\begin{assumption}[Stability] For any $j$, $k$, and $t$,  there exists a function $\mu_{jkt}(\cdot)$ with $\mathbb{E}_{{X}} \{\mu_{jkt}^2(\pmb{X}_{i,t})\}<\infty$ such that, as $n\to \infty$, $\mathbb{E}_{{X}}\{\hat{\mu}_{jkt}(\pmb{X}_{i,t})- {\mu}_{jkt}(\pmb{X}_{i,t})\}^2 {\to} 0$ in probability with respect to the randomness of $\hat{\mu}_{jkt}(\cdot)$ as a function of data, where $\mathbb{E}_{{X}}$ is the expectation with respect to $\pmb{X}_{i,t}$. If $\hat{\mu}_{jkt}(\cdot)$ is not from a finite-dimensional parametric model, then $\hat{\mu}_{jkt}(\cdot)$ and ${\mu}_{jkt}(\cdot)$ also need to satisfy the Donsker condition stated in the Appendix.
\end{assumption}

Theorem \ref{theo: CAN} shows that all weighting and post-stratification estimators are fully robust: they are consistent and asymptotically normal for $\bm\vartheta_{jk}$ under minimal assumptions. 

\begin{theorem} \label{theo: CAN}
    Under Assumption 1 (and Assumption 2 if covariates are adjusted), for fixed $j$ and $k$, and $\star \in \text{\{ipw, sipw, aipw, ps, aps\}}$, $ \sqrt{n}(\hat{\pmb{\vartheta}}_{jk}^{\star} - \pmb{\vartheta}_{jk})$ converges in distribution to the bivariate normal with mean vector 0 and covariance matrix 
    ${\rm Var} ((\phi^\star_{jk} (R_i), \phi^\star_{kj}(R_i) )^T)$ as $n\to\infty$, 
where $R_i$ denotes the observed data from individual $i$. The explicit form of the influence function $\phi^\star_{jk}(R_i)$ is provided for each estimator as follows:

\begin{enumerate}[label=\alph*), leftmargin=*, align=left]
    \item  For the IPW estimator $\hat{\theta}_{jk}^{\text{ipw}}$,
    \begin{align*}
        \phi^{\rm ipw}_{jk}(R_i) = \frac{1}{\sum_{t=1}^T\mathbb{P}(I_{jkt}(i) = 1)}
        \cdot \left \{\sum_{t=1}^T I_{jkt}(i)\left (\frac{I(A_{i,t}=j)Y_{i,t}}{\pi_j^t(Z_{i,t})} -\theta_{jk}\right ) \right\}.
    \end{align*}
    \item For the SIPW estimator $\hat{\theta}_{jk}^{\text{sipw}}$,
    \begin{align*}
        \phi^{\rm sipw}_{jk}(R_i) = \frac{1}{\sum_{t=1}^T\mathbb{P}(I_{jkt}(i) = 1)}\cdot\left \{\sum_{t=1}^T I_{jkt}(i)\left (\frac{I(A_{i,t}=j)(Y_{i,t}-\theta_{jk})}{\pi_j^t(Z_{i,t})} \right ) \right\}.
    \end{align*}
    \item For the AIPW estimator $\hat{\theta}_{jk}^{\text{aipw}}$,
    \begin{align*}
        &\phi^{\rm aipw}_{jk}(R_i) = \frac{1}{\sum_{t=1}^T\mathbb{P}(I_{jkt}(i) = 1)}\cdot\left \{\sum_{t=1}^T I_{jkt}(i)\left (\frac{I(A_{i,t}=j)[Y_{i,t}-\mu_{jkt}(\pmb{X}_{i,t})]}{\pi_j^t(Z_{i,t} )}+ \mu_{jkt}(\pmb{X}_{i,t}) - \theta_{jk} \right ) \right\}.
    \end{align*}
    \item For the PS estimator $\hat{\theta}_{jk}^{\text{ps}}$,
    \begin{align*}
        &\phi^{\rm ps}_{jk}(R_i) = \frac{1}{\sum_{t=1}^T\mathbb{P}(I_{jkt}(i) = 1)} \cdot  \sum_{t=1}^T\Bigg \{ \sum_{h=1}^{H_{jkt}}\Bigg[\frac{I(A_{i,t} = j,G_{i,t} = h)}{\mathbb{P}(A_{i,t} = j|G_{i,t}=h)}\Big[Y_{i,t} - \mathbb{E}(Y_{i,t}|A_{i,t}=j,G_{i,t}=h)\Big]\\
        &+I(G_{i,t}=h)\Big[\mathbb{E}(Y_{i,t}|A_{i,t}=j,G_{i,t}=h)-\theta_{jk}\Big]\Bigg] \Bigg\}.
    \end{align*}
    where $G_{i,t} = h$ if and only if $i \in \mathcal{I}_{jkt}^{(h)}$.
    \item For the APS estimator $\hat{\theta}_{jk}^{\text{aps}}$,
    \begin{align*}
        &\phi^{\rm aps}_{jk}(R_i) = \frac{1}{\sum_{t=1}^T\mathbb{P}(I_{jkt}(i) = 1)}\sum_{t=1}^T\Bigg \{ \sum_{h=1}^{H_{jkt}}\Bigg [\frac{I(A_{i,t} = j,G_{i,t} = h)}{\mathbb{P}(A_{i,t} = j|G_{i,t}=h)} \\
        &\times\Big[Y_{i,t} - \mu_{jkt}(\pmb{X}_{i,t})+\bar{\mu}_{jkt}(h) - \mathbb{E}(Y_{i,t}|A_{i,t}=j,G_{i,t}=h)\Big]\\
        +&I(G_{i,t}=h)\Big[\mathbb{E}(Y_{i,t}|A_{i,t}=j,G_{i,t}=h)+ \mu_{jkt}(\pmb{X}_{i,t})-\bar{\mu}_{jkt}(h)-\theta_{jk}\Big]\Bigg ] \Bigg\},
    \end{align*}
    where $\bar{\mu}_{jkt}(h):= \mathbb{E}[\mu_{jkt}(\pmb{X}_{i,t})|G_{i,t} = h]$.
\end{enumerate}
\end{theorem}


Following Theorem 1, we emphasize that these large-sample guarantees follow from the master protocol trial's randomization scheme (Assumption 1), along with mild regularity conditions when adjusting for covariates. In particular, consistency and asymptotic normality do not rely on correctly specified outcome models or assumptions about the within-individual correlation structure across episodes. Dependence arising from re-enrollment is permitted, and as we show in the next subsection, inference remains valid under within-person clustering when cluster-robust variance estimators are used.

\subsection{Cluster-robust variance estimation}\label{sec::cluster_robust}

We use plug-in variance estimators to estimate the asymptotic variance in Theorem 1. Specifically, we replace $\mathbb{P}(I_{jkt}(i) = 1),\mu_{jkt}(\cdot),\bar{\mu}_{jkt}(h),\text{ and }\mathbb{E}(Y_{i,t}|A_{i,t}=j,G_{i,t}=h) $ with their empirical counterparts $|\mathcal{I}_{jkt}|/n$, $\hat{\mu}_{jkt}(\cdot)$, $\sum_{i\in \mathcal{I}_{jkt}^{(h)}} \hat{\mu}_{jkt}(\pmb{X}_{i,t})/n(\mathcal{I}_{jkt}^{(h)})$ and $\sum_{i\in \mathcal{I}_{jkt}^{(h)}} I(A_{i,t} = j)Y_{i,t}/n_j(\mathcal{I}_{jkt}^{(h)})$ for all $t = 1,\cdots, T$, and replace $\theta_{jk}$ with the estimator $\hat{\theta}_{jk}^{\star}$. This gives the estimated influence function $\hat{\phi}_{jk}^\star(R_i)$ by summing the episode-level contributions within individual $i$. The variance estimator $\widehat{\mathrm{Var}}\!\left(\hat\theta_{jk}^{\star}\right)$ is the sample variance across individuals,
\[
n\widehat{\mathrm{Var}}\!\left(\hat\theta_{jk}^{\star}\right)
=\frac{1}{n-1}\sum_{i=1}^n\Big [\hat\phi_{jk}^\star(R_i) -\bar{\hat\phi}_{jk}^\star \Big]^2,\;\bar{\hat\phi}_{jk}^\star = \frac{1}{n}\sum_{i=1}^n\hat\phi_{jk}^\star(R_i)
\]
Since each $\hat{\phi}_{jk}^\star(R_i)$ aggregates all episode-level contributions for individual $i$, the variance estimator is cluster-robust with clustering at the individual level and remains valid under within-person dependence from re-enrollment.

\section{Simulation Study}\label{sec::simulation}

We conduct a simulation study based on the SIMPLIFY trial design to examine the finite-sample performance of all estimators in Section \ref{sec::estimation}. The total {number of unique participants} $n$ is 600 or 1,200.  The observed covariate vector is $\pmb{W}_{i,t} = (X_{i,(b)}, X_{i,(cat)},X_{it,(c)},X_{i,EW})$, where $X_{i,(b)}$ is an episode-invariant binary covariate with $\mathbb{P}(X_{i,(b)} = 1) = 0.5$, $X_{i,(cat)}$ is an episode-invariant categorical covariate indicating disease subtype, with $\mathbb{P}(X_{i,(cat)} = 0) = 0.03,\mathbb{P}(X_{i,(cat)} = 1) = 0.24,\mathbb{P}(X_{i,(cat)} = 2) = 0.73$. $X_{it,(c)}$ is an episode-varying continuous covariate generated from a log-normal distribution that depends on disease subtype: $X_{i1,(c)}|X_{i,(cat)} = k \sim \log \mathbb{N}(\mu_k,0.4)$, where $\mu_0=3.25, \mu_1 =3.1, \mu_2 = 3.0$. To better mimic the SIMPLIFY trial, $X_{i1,(c)}$ is truncated to the interval $[12,70]$. For re-enrolled participants, $X_{i2,(c)} = X_{i1,(c)} + V_i$, where $V_i\sim \text{Uniform}(0,1)$. $X_{i,EW}$ is an episode-invariant categorical covariate indicating enrollment window, with $\mathbb{P}(X_{i,EW} =1) = 5/6$ and $\mathbb{P}(X_{i,EW} =2) = 1/6$. We also generate an episode-invariant unmeasured variable $U_{i}\sim \mathbb{N}(0,1)$ and random errors $\epsilon_{i,t}^{(\cdot)} \sim \mathbb{N}(0,1)$. The potential outcomes are generated according to $Y_{i,t}^{(\cdot)}=0.5\,X_{i,(b)}+0.1\,X_{it,(c)}+U_i+\beta^{(\cdot)}X_{i,(cat)}+\delta^{(\cdot)}+\epsilon_{i,t}^{(\cdot)}$, where the arm-specific coefficients $\beta^{(\cdot)}$ and $\delta^{(\cdot)}$ are given by: 
\[
\resizebox{0.75\columnwidth}{!}{$
\begin{array}{@{}c|cccccccccc@{}}
(\cdot) & (1) & (2) & (3) & (1,1) & (1,2) & (1,3) & (2,1) & (2,3) & (3,1) & (3,2)\\ \hline
\beta^{(\cdot)}  & 0.2 & -1   & -0.5 & 0.2 & -0.4 & -0.15 & -0.4 & -0.75 & -0.15 & -0.75 \\
\delta^{(\cdot)} & 0   & -2   & 2    & 0   & -1.5 & 1.5   & -1   & 1     & 0.5   & -0.5
\end{array}
$}
\]

 Individuals are first randomized into one of two sub-studies depending on their enrollment window $X_{i,EW}$ and disease subtypes $X_i^{(cat)}$, and then randomized with equal allocation to treatments within the assigned substudy. Substudy $1$ has treatments $1$ and $2$, and substudy $2$ has treatments $1$ and $3$. The assignment probabilities are in Table \ref{tab:example}. Here, $X_{i,(cat)}$ has the same interpretation as $(\mathrm{HS}_i,\mathrm{DA}_i)$ in the SIMPLIFY study (see the table note).

Following the SIMPLIFY trial, we restrict re-enrollment eligibility to participants with $X_{i,(cat)}= 2$. We consider two scenarios for the re-enrollment procedure:
\begin{enumerate}[leftmargin=*, align=left]
    \item A random 58\% of eligible participants are re-enrolled.
    \item Each eligible participant is re-enrolled with probability $\mathbb{P}(\text{participant $i$ re-enrolled}) = \{1+\exp(\gamma +U_{i})\}^{-1}$, where $\gamma = -0.39$ is chosen so that approximately 58\% of eligible participants are re-enrolled.
\end{enumerate}

We consider the following estimators of the treatment effects of 2 versus 1 and 3 versus 1:

\begin{enumerate}[label=(\roman*), leftmargin=*, align=left]
    \item {\bf IPW-based and PS-based estimators using re-enrolled data:} We estimate the linear contrast $\theta_{21} - \theta_{12}$ and $\theta_{31} - \theta_{13}$ using the IPW, SIPW, AIPW, PS and APS estimators described in Section \ref{sec::estimation}. For AIPW and APS, the outcome models $\hat{\mu}_{jkt}$ and $\hat{\mu}_{kjt}$ are obtained by fitting episode-specific linear regressions of $Y_{i,t}$ on $X_{it,(c)}$ and $X_{i,(b)}$, using data from treatment arms $j$ and $k$ in $\mathcal{I}_{jkt}$, respectively. Note that these working models are misspecified because $X_{i,(cat)}$ is omitted. For PS and APS, individuals are stratified by the values of $\pi_j^t(Z_{i,t})$ and $\pi_k^t(Z_{i,t})$ for each episode $t$. Based on the assignment probabilities in Table \ref{tab:example}, the strata are defined as shown in the Appendix.
    \item {\bf IPW-based and PS-based estimators using episode 1 data only:} We use only observations from episode 1 and apply the same estimators described in Section \ref{sec::estimation} with $T = 1$. This class of estimators was proposed by \citet{qian2025}. 
    \item {\bf Substudy-specific estimators using re-enrolled data:} For comparison, we also evaluate traditional ANOVA, ANCOVA, and ANHECOVA estimators (using linear working models to adjust for $X_{it,(c)}$ and $X_{i,(b)}$) \cite{robincar,Ye02102023} based solely on data from each substudy. This class of estimators was used as the primary analysis in \citet{MayerHamblett2023SIMPLIFY}. We also applied ANHECOVA with linear working models fitted separately for each episode; because the results are similar to those from models fitted on pooled episode data, they are omitted.
    
\end{enumerate}

Note that the episode-1-only and substudy-specific methods target different estimands than ours: the episode-1-only method targets the average treatment effect for episode 1, while the substudy-specific estimators target treatment effects within each substudy. Nevertheless, we can still compare their variances. Moreover, these methods do not involve dependence issues, since each participant appears at most once in episode 1 or in each substudy; thus, standard variance estimators can be directly applied.

Table \ref{Result:scenario1} shows the simulation results across 5,000 simulation runs with $n=600$.  The results contain the average bias, standard deviation (SD), the average of the standard error (SE), and the coverage probability (CP) of the $95\%$ confidence interval. The true values of estimands are calculated analytically. The following summarizes the key findings, which also hold for $n$  =1,200 (see Appendix).

The simulation results support our asymptotic theory. 
All proposed estimators using re-enrolled data exhibit negligible bias. Different re-enrollment mechanisms do not bias the estimators, even when re-enrollment depends on unmeasured factors, because the consistency and asymptotic normality of the proposed estimators depend only on randomization of treatment within each episode. Comparing standard deviations, SIPW improves upon IPW, and covariate adjustment yields clear efficiency gains despite working-model misspecification. Estimated SEs closely match empirical SDs, and CPs are near the nominal 95\% level, demonstrating the validity of the proposed cluster-robust variance estimators.

Although the other two classes of estimators target different estimands, the proposed estimators that use data from multiple episodes achieve smaller SDs than those using only episode 1, corresponding to approximately 20\% variance reduction. The three substudy-specific estimators (ANOVA, ANCOVA, and ANHECOVA) analyze each substudy separately and leverage the re-enrolled data. However, this class of estimators still has larger SDs than the proposed estimators because they do not pool treatment 1 data across the two substudies. Specifically, the proposed SIPW and AIPW achieve approximately 20\% variance reduction compared with substudy-specific ANOVA and ANHECOVA, respectively.

\begin{table}[htbp]
\centering
\caption{Simulation results based on 5,000 simulation runs ($n=600$). The estimand for the proposed estimators is $\theta_{jk} - \theta_{kj}$; for the episode 1 only estimators is $\theta_{jkt} - \theta_{kjt}$ with $t=1$; and for the substudy-specific methods is the substudy treatment effect.}
\label{Result:scenario1}
\resizebox{.9\columnwidth}{!}{%
\begin{tabular}{llcccccccc}
\toprule
&
 & \multicolumn{4}{c}{Treatment 2 vs 1} & \multicolumn{4}{c}{Treatment 3 vs 1}\\
\cmidrule(lr){3-6}\cmidrule(lr){7-10}
\multicolumn{2}{l}{\bf Re-enrollment scenario 1} & \multicolumn{4}{c}{Truth $ = -3.928$} & \multicolumn{4}{c}{Truth $ = 0.826$}\\
\cmidrule(lr){3-7}\cmidrule(lr){7-10}
Class of methods &Estimator&Bias&SD&SE&CP&Bias&SD&SE&CP\\
\midrule
Proposed & IPW & -0.004 & 0.206 & 0.206 & 0.952 & 0.010 & 0.338 & 0.333 & 0.945\\
 & SIPW & -0.003 & 0.171 & 0.171 & 0.951 & -0.001 & 0.158 & 0.158 & 0.948\\
 & AIPW & -0.002 & 0.143 & 0.140 & 0.945 & 0.002 & 0.134 & 0.132 & 0.945\\
 & PS & -0.005 & 0.155 & 0.154 & 0.950 & -0.004 & 0.144 & 0.143 & 0.950\\
 & APS & -0.003 & 0.129 & 0.126 & 0.948 & -0.001 & 0.121 & 0.118 & 0.941\\
\hline
\multicolumn{2}{c}{} & \multicolumn{4}{c}{Truth $ = -4.305$} & \multicolumn{4}{c}{Truth $ = 0.773$} \\
\addlinespace[2pt]
\hline
Episode 1 only & IPW & -0.004 & 0.228 & 0.227 & 0.951 & 0.012 & 0.401 & 0.391 & 0.946\\
 & SIPW & -0.005 & 0.188 & 0.190 & 0.951 & 0.000 & 0.174 & 0.175 & 0.948\\
 & AIPW & -0.002 & 0.161 & 0.159 & 0.946 & 0.003 & 0.149 & 0.147 & 0.946\\
 & PS & -0.006 & 0.174 & 0.175 & 0.951 & -0.003 & 0.158 & 0.158 & 0.950\\
 & APS & -0.003 & 0.144 & 0.141 & 0.943 & 0.000 & 0.131 & 0.128 & 0.943\\
\hline
\multicolumn{2}{c}{} & \multicolumn{4}{c}{Truth $= -3.673$} & \multicolumn{4}{c}{Truth $= 0.937$} \\
\hline
\addlinespace[2pt]
Substudy specific & ANOVA & -0.002 & 0.188 & 0.188 & 0.951 & -0.002 & 0.168 & 0.166 & 0.946\\
 & ANCOVA & -0.002 & 0.161 & 0.159 & 0.946 & 0.000 & 0.142 & 0.141 & 0.944\\
 & ANHECOVA & -0.002 & 0.161 & 0.159 & 0.946 & 0.000 & 0.142 & 0.141 & 0.944\\
\midrule \midrule
\multicolumn{2}{l}{\bf Re-enrollment scenario 2} & \multicolumn{4}{c}{Truth $ = -3.928$} & \multicolumn{4}{c}{Truth $ = 0.826$}\\
\cmidrule(lr){3-7}\cmidrule(lr){7-10}
Class of methods &Estimator&Bias&SD&SE&CP&Bias&SD&SE&CP\\
\midrule
Proposed & IPW & 0.008 & 0.198 & 0.202 & 0.951 & 0.000 & 0.335 & 0.331 & 0.945\\
 & SIPW & 0.008 & 0.167 & 0.168 & 0.951 & -0.005 & 0.160 & 0.160 & 0.949\\
 & AIPW & 0.012 & 0.143 & 0.139 & 0.941 & -0.004 & 0.135 & 0.131 & 0.943\\
 & PS & 0.006 & 0.153 & 0.153 & 0.948 & -0.006 & 0.145 & 0.143 & 0.947\\
 & APS & 0.012 & 0.128 & 0.125 & 0.941 & -0.004 & 0.121 & 0.118 & 0.942\\
\hline
\multicolumn{2}{c}{} & \multicolumn{4}{c}{Truth $= -4.305$} & \multicolumn{4}{c}{Truth $=  0.773$} \\
\addlinespace[2pt]
\hline
Episode 1 only  & IPW & -0.010 & 0.225 & 0.227 & 0.952 & 0.004 & 0.396 & 0.391 & 0.948\\
 & SIPW & -0.012 & 0.191 & 0.190 & 0.946 & -0.006 & 0.173 & 0.175 & 0.951\\
 & AIPW & -0.005 & 0.163 & 0.159 & 0.942 & -0.003 & 0.148 & 0.147 & 0.948\\
 & PS & -0.012 & 0.175 & 0.174 & 0.944 & -0.005 & 0.160 & 0.158 & 0.947\\
 & APS & -0.005 & 0.144 & 0.141 & 0.945 & -0.003 & 0.130 & 0.128 & 0.946\\
\hline
\multicolumn{2}{c}{} & \multicolumn{4}{c}{Truth $= -3.673$} & \multicolumn{4}{c}{Truth $= 0.937$} \\
\hline
\addlinespace[2pt]
Substudy specific  & ANOVA & 0.014 & 0.181 & 0.184 & 0.952 & -0.001 & 0.171 & 0.170 & 0.948\\
 & ANCOVA & 0.017 & 0.156 & 0.155 & 0.945 & -0.001 & 0.146 & 0.144 & 0.946\\
 & ANHECOVA & 0.017 & 0.156 & 0.155 & 0.945 & -0.001 & 0.146 & 0.144 & 0.947\\
\bottomrule
\addlinespace[5pt]
\end{tabular}}
\end{table}

\section{Data Application: SIMPLIFY Trial}\label{sec::real}

  As described in Section~\ref{sec::setup_simplify}, SIMPLIFY evaluates three {treatment regimens after establishment of ETI}: continue existing therapies (treatment 1), discontinue HS (treatment 2), and discontinue DA (treatment 3). For the comparison of treatment 2 versus treatment 1, the episode-1 ECE population includes all individuals on HS or HS+DA at baseline, while the episode-2 ECE population includes all individuals on HS+DA at baseline who entered the DA substudy in episode 1 and were re-enrolled in episode 2. Similarly, {for the comparison of treatment 3 versus treatment 1, the episode-1 ECE population includes all individuals on DA or HS+DA at baseline, while the episode-2 ECE population includes all individuals on HS+DA at baseline who entered the HS substudy in episode 1 and were re-enrolled in episode 2} (Figure~\ref{fig:SIMPLIFY_illustration}). These populations illustrate the general principle that the episode-$t$ ECE population includes all individuals who could have been randomized to both treatments being compared at episode $t$. We excluded 10
(1.7\%) participants due to missing outcomes at episode $1$, resulting in a sample size of $n = 584$ and a total of 838 observations.

 For the defined estimands $\theta_{jk}$ versus $ \theta_{kj}$, we assess the IPW, SIPW, AIPW, PS, and APS estimators. The covariate vector $Z_{i,t}$, which determines treatment assignment at episode $t$, includes indicators for baseline HS/DA use, enrollment window, and prior substudy enrollment. For the AIPW and APS estimators, $\hat{\mu}_{jkt}(\cdot)$ is obtained by fitting an episode-specific linear regression of $Y_{i,t}$ on covariates $\pmb{X}_{i,t}$ using observations in the treatment arm $j$ within $\mathcal{I}_{jkt}$ at episode $t$. The covariates $\pmb{X}_{i,t}$ include $Z_{i,t}$ along with additional variables: two continuous covariates (age and $\text{ppFEV}_1$) measured at episode-$t$ baseline; three binary baseline covariates (sex, race [white vs. non-white], and ethnicity [Hispanic or Latino vs. not Hispanic or Latino]); and one three-level baseline covariate (genotype: delta F508 homozygous, heterozygous, or other/unknown). As in the simulation study, we also examine the same five estimators using only episode-$1$ or episode-$2$ data, and apply ANOVA, ANCOVA, and ANHECOVA to all data (including re-enrolled observations) within each substudy. A comparison of these approaches is 
 summarized in Table \ref{tab:comparison}. 

\begin{figure}[htbp] 
    \centering
    \includegraphics[width=\columnwidth, trim=1.5cm 0cm 0cm 0cm, clip]{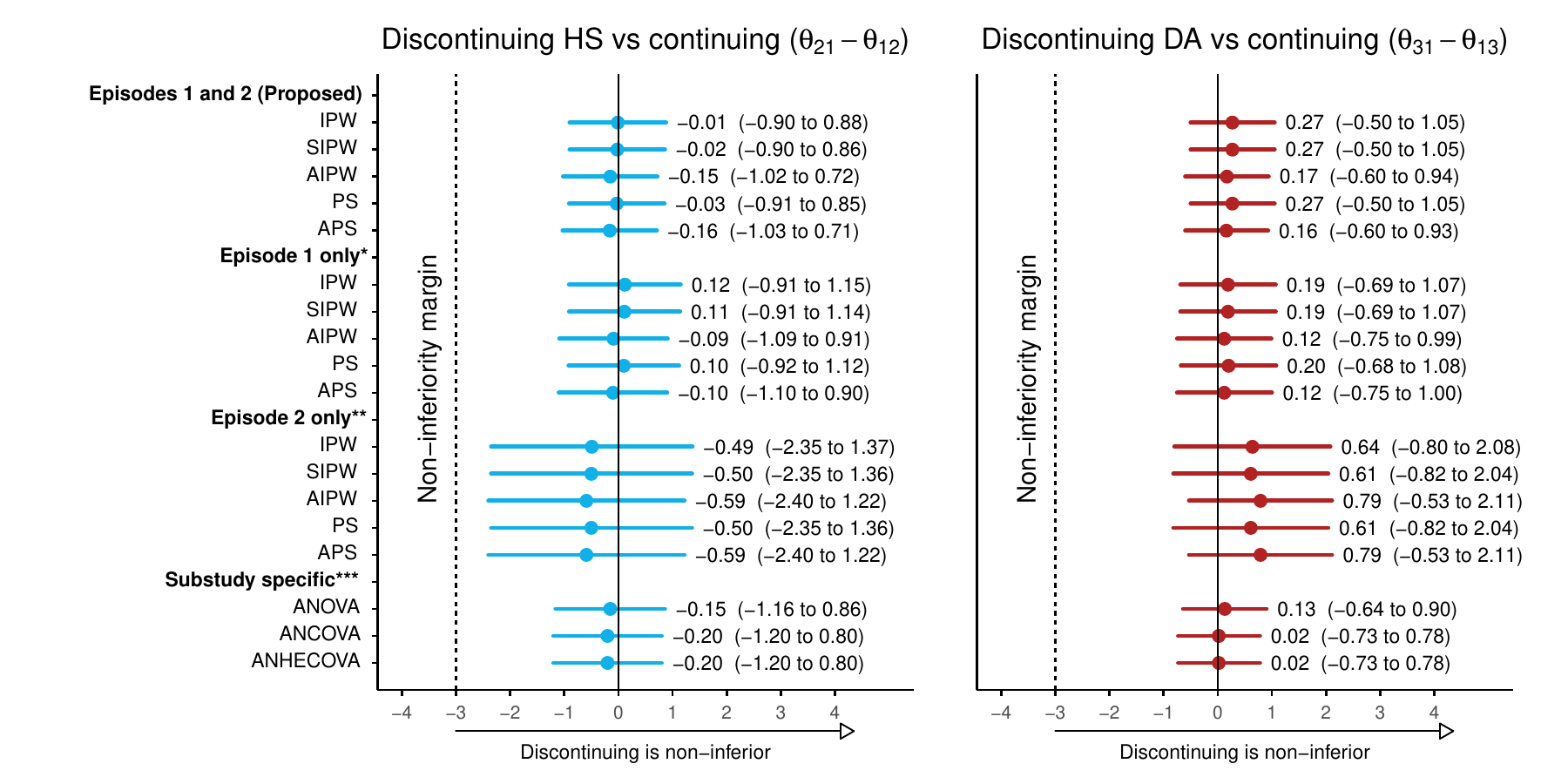}
    \caption{Estimates and 95\% confidence intervals (in \%) in the SIMPLIFY trial.}
    \label{fig:SIMPLIFY_forest}
    \begin{flushleft}\footnotesize
    \textsuperscript{*}:\;The corresponding estimand is $\theta_{jkt}-\theta_{kjt}, t = 1$.\\
    \textsuperscript{**}:\;The corresponding estimand is $\theta_{jkt}-\theta_{kjt}, t = 2$.\\
    \textsuperscript{***}:\;The corresponding estimand is the average treatment effect within the substudy.
  \end{flushleft}
\end{figure}

\begin{table}[!htbp]
\centering
\caption{Comparison of three estimation approaches}
\label{tab:comparison}
\small
\setlength{\tabcolsep}{5pt}
\renewcommand{\arraystretch}{1.15}
\resizebox{\textwidth}{!}{
\begin{tabular}{L{0.12\textwidth} C{0.22\textwidth} C{0.15\textwidth} C{0.15\textwidth} L{0.4\textwidth}}
\toprule
\textbf{Method}
& \textbf{Estimand}
& \textbf{\shortstack[c]{Uses re-enrolled\\data}}
& \textbf{\shortstack[c]{Uses shared\\controls}}
& \textbf{Notes on estimand} \\
\midrule
Proposed
& $\pmb{\vartheta}_{jk}$
& $\checkmark$
& $\checkmark$
& 
Per-episode added effect defined on episode-$t$ ECE population across all episodes; does not depend on randomization ratios or platform trial format. \\
\addlinespace[6pt]
Qian et al.\ (2025)
& $\pmb{\vartheta}_{jk1}$
& $\times$
& $\checkmark$
& 
Episode-1 effect defined on episode-1 ECE population; does not depend on randomization ratios or platform trial format. \\
\addlinespace[6pt]
Substudy-specific
& {\footnotesize$
\left(
\begin{array}{c}
\mathbb{E}(Y_{i,t}\mid A_{i,t}=j,S_{i,t}=s)\\
\mathbb{E}(Y_{i,t}\mid A_{i,t}=k,S_{i,t}=s)
\end{array}
\right)$}
& $\checkmark$
& $\times$
& 
Defined on population assigned to substudy $s$; depends on randomization ratios or platform trial format, complicating interpretation. \\
\bottomrule
\end{tabular}
}
\end{table}

 The point estimates and 95\% confidence intervals are shown in Figure~\ref{fig:SIMPLIFY_forest}. For all five estimators (IPW, SIPW, AIPW, PS, and APS), point estimates are similar when targeting the same estimand defined on the ECE population (i.e., $\theta_{jk} - \theta_{kj}$ for episodes 1 and 2 pooled; $\theta_{jkt}-\theta_{kjt}$ for episode $t$ only, $t=1,2$). Since all confidence intervals lie to the right of the pre-defined non-inferiority margin of $-3\%$ (the vertical dashed line in Figure~\ref{fig:SIMPLIFY_forest}), non-inferiority of discontinuing therapy can be clearly established for both HS and DA, consistent with the findings reported in \citet{MayerHamblett2023SIMPLIFY}.

 The point estimates for the episode-2 contrast ($\theta_{jk2} - \theta_{kj2}$), estimated using episode-2 data only, differ slightly from the episode-1 contrast but still support non-inferiority. 
 This difference likely reflects the nature of re-enrollment in master protocol trials: re-enrolled participants may differ systematically from those who were not re-enrolled, and carryover effects from the previous episode may also contribute. The proposed method uses data from both episodes and yields narrower confidence intervals \textcolor{black}{(22\%-26\% variance reduction)} than using episode-1 data alone.
 
For comparison, we also include the ANOVA, ANCOVA, and ANHECOVA estimators based solely on data from each individual substudy. These estimators also use re-enrolled data but target the analysis populations assigned to either the HS or DA substudy, so their interpretations differ from those of the five proposed estimators and they depend on the randomization ratios or the platform trial format. Moreover, the ANOVA, ANCOVA, and ANHECOVA estimators yield wider confidence intervals than the five proposed estimators when pooling data across episodes. This occurs because, under the five proposed estimators, participants in the HS+DA stratum who are assigned to treatment 1 contribute to both the 2 vs. 1 and 3 vs. 1 comparisons, serving as a shared control. In contrast, the substudy-specific estimators use only the subset of these shared control {participants} within each corresponding substudy.

\section{Discussion} \label{sec::discuss}

In this paper, we develop a statistical framework for master protocol trials with re-enrollment and construct robust, efficient estimators that fully utilize information both across substudies and across episodes for individuals who re-enroll. We first establish a statistical framework using potential outcomes and rigorous assumptions that characterize the randomization procedure under re-enrollment. Our framework is general as it can accommodate all types of master protocol trials (including umbrella, basket, and platform) and trials with more than two episodes. We then define clinically meaningful estimands and construct consistent, asymptotically normal estimators that are valid under the trial's randomization mechanism. Importantly, these estimators are unbiased even under selective re-enrollment, which commonly occurs in practice. 

Specifically, we define the episode-$t$ ECE population for episode-specific treatment effects and introduce the per-episode added-effect estimand as our primary estimand, which is a weighted average of episode-specific treatment effects with weights proportional to the probability of belonging to each episode's ECE population. We propose weighting, stratification, and covariate adjustment methods to construct robust estimators that take full advantage of data from re-enrolled participants and across substudies, and derive cluster-robust variance estimators that account for within-participant correlation induced by re-enrollment. In principle, additional efficiency gains may be achievable by optimizing the weights for combining episode-specific effects based on an estimated within-participant dependence structure. However, such approaches can yield estimands that are less interpretable, so we do not pursue them here.

Re-analysis of the SIMPLIFY trial demonstrates that the proposed methods leveraging re-enrolled data yield approximately 25\% variance reduction compared to estimators {of \citet{qian2025}} using only episode-1 data. This highlights that for trials permitting re-enrollment, restricting analysis to the first episode discards valuable information. Moreover, the conventional approach of analyzing each substudy separately is not only less efficient, because it forgoes sharing control data across substudies, but also targets an estimand that depends on the randomization ratio and is therefore difficult to interpret.

Our findings suggest several practical considerations for designing master protocol trials that permit re-enrollment. Beyond the statistical efficiency gains demonstrated in this paper, re-enrollment offers additional operational advantages: it accelerates trial enrollment by avoiding repeated screening procedures and may improve protocol compliance, as re-enrolling participants are already familiar with study procedures and have demonstrated commitment to the trial. Re-enrollment also creates scientific opportunities: it enables investigation of carryover effects and allows researchers to study treatment effects of sequential or combination therapies. These questions cannot be addressed in single-episode designs and warrant further statistical development.

\bibliographystyle{unsrtnat}
\bibliography{references-arxiv.bib}

\newpage
\begin{center}
{\LARGE \bfseries Supplementary Materials for ``Evolving Longitudinal Patient Histories and Re-enrollment in Master Protocol Trials''}
\end{center}
\vspace{2em}

\setcounter{section}{0}
\renewcommand{\thesection}{S\arabic{section}}

\section*{Web Appendix A.}
\label{sec::weba}

\subsection*{PS/APS strata definition}

As noted in Section 4, Table~\ref{sim:PS_strata} defines the PS and APS strata, and for $T=1$, we restrict to the first three strata.

\begin{table}[htbp]
\centering
\caption{Details of strata in the simulation study}
\begin{tabular}{cccccc}
\toprule
Estimand & Stratum & $j$ & $k$ & $t$ &  Stratum Definition \\
\midrule
\multirow{5}{*}{$\theta_{21}-\theta_{12}$}
  & $\mathcal{I}^{(1)}_{jkt}$ & 1&2&1& $\{\, i : \pi_{j}^{t}(i)=0.5,\ \pi_{k}^{t}(i)=0.5 \,\}$ \\ [0.6ex]
  & $\mathcal{I}^{(2)}_{jkt}$ & 1&2&1& $\{\, i : \pi_{j}^{t}(i)=0.5,\ \pi_{k}^{t}(i)=0.25 \,\}$ \\ [0.6ex]
  & $\mathcal{I}^{(3)}_{jkt}$ & 1&2&1& $\{\, i : \pi_{j}^{t}(i)=0.5,\ \pi_{k}^{t}(i)=0.375 \,\}$  \\ [0.6ex]
  & $\mathcal{I}^{(1)}_{jkt}$ & 1&2&2&  $\{\, i : \pi_{j}^{t}(i)=0.5,\ \pi_{k}^{t}(i)=0.5 \,\}$  \\
\midrule
\multirow{5}{*}{$\theta_{31}-\theta_{13}$}
  & $\mathcal{I}^{(1)}_{jkt}$ & 1&3&1&  $\{\, i : \pi_{j}^{t}(i)=0.5,\ \pi_{k}^{t}(i)=0.5 \,\}$ \\ [0.6ex]
  & $\mathcal{I}^{(2)}_{jkt}$ & 1&3&1&  $\{\, i : \pi_{j}^{t}(i)=0.5,\ \pi_{k}^{t}(i)=0.25 \,\}$ \\ [0.6ex]
  & $\mathcal{I}^{(3)}_{jkt}$ & 1&3&1&  $\{\, i : \pi_{j}^{t}(i)=0.5,\ \pi_{k}^{t}(i)=0.125 \,\}$  \\ [0.6ex]
  & $\mathcal{I}^{(1)}_{jkt}$ & 1&3&2&  $\{\, i : \pi_{j}^{t}(i)=0.5,\ \pi_{k}^{t}(i)=0.5 \,\}$  \\
\bottomrule
\end{tabular}
\vspace{2mm}
\label{sim:PS_strata}
\end{table}

\section*{Web Appendix B.}
\label{sec::webb}

\subsection*{Additional Simulation results when $n=1{,}200$}

The additional simulation mirrors Section 4 but increases the sample size to $n=1,200$. The main conclusions persist: SIPW has lower SD than IPW; covariate adjustment yields efficiency gains even when the working model is misspecified; and estimators that treat each substudy as a stand-alone trial have larger SDs than their counterparts defined on ECE populations. All estimators also exhibit smaller standard errors than in the smaller-sample setting.

\begin{table}[htbp]
\centering
\caption{Simulation results based on 5,000 simulation runs ($n=1,200$). The estimand for the proposed estimators is $\theta_{jk} - \theta_{kj}$; for the episode 1 only estimators is $\theta_{jkt} - \theta_{kjt}$ with $t=1$; and for the substudy-specific methods is the substudy treatment effect.}
\label{Result:scenario1-largen}
\resizebox{.9\columnwidth}{!}{%
\begin{tabular}{llcccccccc}
\toprule
&
 & \multicolumn{4}{c}{Treatment 2 vs 1} & \multicolumn{4}{c}{Treatment 3 vs 1}\\
\cmidrule(lr){3-6}\cmidrule(lr){7-10}
\multicolumn{2}{l}{\bf Re-enrollment scenario 1} & \multicolumn{4}{c}{Truth $ = -3.928$} & \multicolumn{4}{c}{Truth $ = 0.826$}\\
\cmidrule(lr){3-7}\cmidrule(lr){7-10}
Class of methods &Estimator&Bias&SD&SE&CP&Bias&SD&SE&CP\\
\midrule
Proposed & IPW & -0.003 & 0.146 & 0.146 & 0.951 & 0.007 & 0.243 & 0.235 & 0.944\\
 & SIPW & -0.003 & 0.119 & 0.121 & 0.953 & -0.003 & 0.112 & 0.112 & 0.949\\
 & AIPW & -0.004 & 0.099 & 0.099 & 0.950 & 0.000 & 0.093 & 0.093 & 0.950\\
 & PS & -0.005 & 0.107 & 0.109 & 0.952 & -0.005 & 0.101 & 0.102 & 0.950\\
 & APS & -0.005 & 0.088 & 0.089 & 0.953 & -0.002 & 0.084 & 0.084 & 0.945\\
\hline
\multicolumn{2}{c}{} & \multicolumn{4}{c}{Truth $ = -4.305$} & \multicolumn{4}{c}{Truth $ = 0.773$} \\
\addlinespace[2pt]
\hline
Episode 1 only & IPW & -0.004 & 0.162 & 0.161 & 0.949 & 0.008 & 0.290 & 0.277 & 0.939\\
 & SIPW & -0.005 & 0.131 & 0.135 & 0.953 & -0.002 & 0.123 & 0.124 & 0.950\\
 & AIPW & -0.005 & 0.113 & 0.113 & 0.949 & 0.001 & 0.104 & 0.104 & 0.943\\
 & PS & -0.006 & 0.120 & 0.124 & 0.954 & -0.005 & 0.112 & 0.112 & 0.952\\
 & APS & -0.006 & 0.100 & 0.100 & 0.953 & -0.002 & 0.091 & 0.091 & 0.948\\
\hline
\multicolumn{2}{c}{} & \multicolumn{4}{c}{Truth $= -3.673$} & \multicolumn{4}{c}{Truth $= 0.937$} \\
\hline
\addlinespace[2pt]
Substudy specific & ANOVA & -0.002 & 0.132 & 0.133 & 0.953 & -0.004 & 0.119 & 0.117 & 0.949\\
 & ANCOVA & -0.003 & 0.113 & 0.112 & 0.948 & -0.002 & 0.100 & 0.100 & 0.954\\
 & ANHECOVA & -0.003 & 0.113 & 0.112 & 0.948 & -0.002 & 0.100 & 0.100 & 0.954\\
\midrule \midrule
\multicolumn{2}{l}{\bf Re-enrollment scenario 2} & \multicolumn{4}{c}{Truth $ = -3.928$} & \multicolumn{4}{c}{Truth $ = 0.826$}\\
\cmidrule(lr){3-7}\cmidrule(lr){7-10}
Class of methods &Estimator&Bias&SD&SE&CP&Bias&SD&SE&CP\\
\midrule
Proposed & IPW & 0.008 & 0.144 & 0.143 & 0.949 & -0.002 & 0.241 & 0.234 & 0.943\\
 & SIPW & 0.008 & 0.118 & 0.119 & 0.951 & -0.005 & 0.112 & 0.114 & 0.954\\
 & AIPW & 0.011 & 0.099 & 0.099 & 0.951 & -0.003 & 0.093 & 0.093 & 0.951\\
 & PS & 0.007 & 0.106 & 0.108 & 0.955 & -0.006 & 0.102 & 0.102 & 0.952\\
 & APS & 0.011 & 0.087 & 0.089 & 0.953 & -0.003 & 0.084 & 0.083 & 0.948\\
\hline
\multicolumn{2}{c}{} & \multicolumn{4}{c}{Truth $= -4.305$} & \multicolumn{4}{c}{Truth $=  0.773$} \\
\addlinespace[2pt]
\hline
Episode 1 only  & IPW & -0.010 & 0.163 & 0.161 & 0.946 & 0.003 & 0.286 & 0.276 & 0.941\\
 & SIPW & -0.011 & 0.134 & 0.134 & 0.946 & -0.004 & 0.122 & 0.124 & 0.955\\
 & AIPW & -0.006 & 0.114 & 0.113 & 0.949 & -0.001 & 0.103 & 0.104 & 0.954\\
 & PS & -0.011 & 0.122 & 0.124 & 0.949 & -0.004 & 0.113 & 0.112 & 0.950\\
 & APS & -0.006 & 0.100 & 0.100 & 0.951 & -0.001 & 0.090 & 0.091 & 0.950\\
\hline
\multicolumn{2}{c}{} & \multicolumn{4}{c}{Truth $= -3.673$} & \multicolumn{4}{c}{Truth $= 0.937$} \\
\hline
\addlinespace[2pt]
Substudy specific  & ANOVA & 0.013 & 0.128 & 0.130 & 0.952 & -0.002 & 0.121 & 0.120 & 0.947\\
 & ANCOVA & 0.016 & 0.108 & 0.110 & 0.951 & -0.001 & 0.102 & 0.102 & 0.950\\
 & ANHECOVA & 0.016 & 0.108 & 0.110 & 0.951 & -0.001 & 0.102 & 0.102 & 0.950\\
\bottomrule
\addlinespace[5pt]
\end{tabular}}
\end{table}

\newpage
\section*{Web Appendix C.}
\label{sec::webc}

\subsection*{Proof of proposition 1}

\textbf{Proposition 1: } The IPW identification formula of $\theta_{jkt}$ at each episode $t$:
\begin{align*}
    \mathbb{E}\Big [ \frac{I(A_{i,t} = j)\cdot Y_{i,t}}{\pi_{j}^t(Z_{i,t})}|I_{jkt}(i) = 1 \Big] = \mathbb{E}\{Y_{i,t}^{(\bar{\pmb{A}}_{i,t-1},j)}|i\in \mathcal{I}_{jkt}\}=\theta_{jkt}
\end{align*}

\textit{Proof:}

Since $I_{jkt}(i) = 0$ when $T_i < t$, we have:

\begin{align*}
    &\mathbb{E}\Big[ \frac{I(A_{i,t} = j)\cdot Y_{i,t}}{\pi_{j}^t(Z_{i,t})}|I_{jkt}(i) = 1\Big] = \mathbb{E}\Big[ \frac{I_{jkt}(i)\cdot I(A_{i,t} = j)\cdot Y_{i,t}}{\pi_{j}^t(Z_{i,t})}|I_{jkt}(i) = 1\Big]\\
    = &\mathbb{E}\Big[ \frac{I_{jkt}(i)\cdot I(A_{i,t} = j)\cdot Y_{i,t}}{\pi_{j}^t(Z_{i,t})}\Big]/\mathbb{P}(I_{jkt}(i) = 1)\\
    = & \mathbb{E}\left\{ \mathbb{E}\left[\dfrac{I_{jkt}(i)\cdot \sum_{\bar{a}_{t-1} \in \mathcal{A}_{t-1}}I(\bar{\pmb{A}}_{i,t-1}=\bar{a}_{t-1}) I(A_{i,t} = j)Y_{i,t}^{(\bar{a}_{t-1},j)}}{\pi_{j}^t(Z_{i,t})}\Big|\mathcal{F}_{i,t-1},I(T_i \geq t)\right]\right \}/{\mathbb{P}(I_{jkt}(i) = 1)}\\
    =& \mathbb{E}\Bigg\{I(T_i \geq t)\cdot \mathbb{E}\Bigg[ I_{jkt}(i)\cdot \sum_{\bar{a}_{t-1} \in \mathcal{A}_{t-1}}I(\bar{\pmb{A}}_{i,t-1}=\bar{a}_{t-1})Y_{i,t}^{(\bar{a}_{t-1},j)}\Big | \mathcal{F}_{i,t-1},T_i \geq t \Bigg] \Bigg\} /{\mathbb{P}(I_{jkt}(i) = 1)}\\
    =& \mathbb{E}\Bigg\{\mathbb{E}\Bigg[ I_{jkt}(i)\cdot\sum_{\bar{a}_{t-1} \in \mathcal{A}_{t-1}}I(\bar{\pmb{A}}_{i,t-1}=\bar{a}_{t-1})Y_{i,t}^{(\bar{a}_{t-1},j)}\Big | \mathcal{F}_{i,t-1},I(T_i \geq t) \Bigg] \Bigg\} /{\mathbb{P}(I_{jkt}(i) = 1)}\\
    =& \frac{\sum_{\bar{a}_{t-1} \in \mathcal{A}_{t-1}}\mathbb{E}[I_{jkt}(i)\cdot I(\bar{\pmb{A}}_{i,t-1}=\bar{a}_{t-1})Y_{i,t}^{(\bar{a}_{t-1},j)}]}{\mathbb{P}(I_{jkt}(i) = 1)} \\
    =& \sum_{\bar{a}_{t-1} \in \mathcal{A}_{t-1}} \mathbb{E}[I(\bar{\pmb{A}}_{i,t-1}=\bar{a}_{t-1})Y_{i,t}^{(\bar{a}_{t-1},j)}|i\in \mathcal{I}_{jkt}] \\
    =& \mathbb{E}\{Y_{i,t}^{(\bar{\pmb{A}}_{i,t-1},j)}|i\in \mathcal{I}_{jkt}\} \\
    =& \theta_{jkt}
\end{align*}

The third equality is based on Assumption 1: $A_{i,t} \perp Y_{i,t}^{(\bar{a}_{t-1},j)}| \mathcal{F}_{i,t-1},T_i \geq t$ and $\mathbb{P}(A_{i,t} = j|\mathcal{F}_{i,t-1},T_i\geq t)  = \mathbb{P}(A_{i,t} = j|Z_{i,t},T_i\geq t) =\pi_{j}^{t}(Z_{i,t})$.

\subsection*{Proof of proposition 2}

\textbf{Proposition 2: } The PS identification formula of $\theta_{jkt}$ at each episode $t$:
\begin{align*}
    \sum_{h=1}^{H_{jkt}}\mathbb{P}[G_{i,t}=h|I_{jkt}(i)=1]\cdot \mathbb{E}[Y_{i,t}|A_{i,t}=j,G_{i,t} = h]= \mathbb{E}\{Y_{i,t}^{(\bar{\pmb{A}}_{i,t-1},j)}|i\in \mathcal{I}_{jkt}\}=\theta_{jkt}.
\end{align*}

\textit{Proof:}

From Assumption 1 and the property of the propensity score, we have that within the episode-$t$ ECE population, $A_{i,t}\perp Y_{i,t}^{(\bar{\pmb{A}}_{i,t-1},j)}|\allowbreak\pi_{j}^t(Z_{i,t})$, and thus $A_{i,t}\perp Y_{i,t}^{(\bar{\pmb{A}}_{i,t-1},j)}|\pi_{j}^t(Z_{i,t}),G_{i,t}$. Since $\pi_{j}^t(Z_{i,t})$ is constant within each stratum defined by $G_{i,t}$, we have $A_{i,t}\perp Y_{i,t}^{(\bar{\pmb{A}}_{i,t-1},j)} | G_{i,t}$.

For episode $t$, the stratum indicator $G_{i,t}\in\{1,\ldots,H_{jkt}\}$ is defined only for individuals in the episode-$t$ ECE population; in particular, $G_{i,t}=h$ implies $I_{jkt}(i)=1$. Hence conditioning on $G_{i,t}$ implicitly conditions on $I_{jkt}(i)=1$. Therefore, based on the assumption of consistency:

\begin{align*}
    &\sum_{h=1}^{H_{jkt}}\mathbb{P}[G_{i,t}=h|I_{jkt}(i)=1]\cdot \mathbb{E}[Y_{i,t}|A_{i,t}=j,G_{i,t} = h] \\
    =& \sum_{h=1}^{H_{jkt}}\mathbb{P}[G_{i,t}=h|I_{jkt}(i)=1]\cdot \mathbb{E}[Y_{i,t}^{(\bar{\pmb{A}}_{i,t-1},j)}|A_{i,t}=j,G_{i,t} = h]\\
    =& \sum_{h=1}^{H_{jkt}}\mathbb{P}[G_{i,t}=h|I_{jkt}(i)=1]\cdot \mathbb{E}[Y_{i,t}^{(\bar{\pmb{A}}_{i,t-1},j)}|G_{i,t} = h]\\
    =& \sum_{h=1}^{H_{jkt}}\mathbb{P}[G_{i,t}=h|I_{jkt}(i)=1]\cdot \mathbb{E}[Y_{i,t}^{(\bar{\pmb{A}}_{i,t-1},j)}|G_{i,t} = h,I_{jkt}(i) = 1]\\
    =& \mathbb{E}[Y_{i,t}^{(\bar{\pmb{A}}_{i,t-1},j)}|I_{jkt}(i) = 1] \\
    =& \theta_{jkt}
\end{align*}

\subsection*{Proof of Theorem 1 a)}

\textbf{Theorem 1 a): } Asymptotic distribution of the IPW estimator $\hat{\theta}_{jk}^{\text{ipw}}$.

Under assumptions stated in Theorem 1,
\begin{align*}
    \sqrt{n}(\hat{\theta}_{jk}^{\text{ipw}}-\theta_{jk}) \overset{d}{\to}\mathbb{N}(0,\text{Var}[ \phi_{jk}^{\text{ipw}}(R_i)]),
\end{align*}

with the IF: $\phi^{\text{ipw}}_{jk}(R_i) = \frac{1}{\sum_{t=1}^T\mathbb{P}(I_{jkt}(i) = 1)}\cdot  \left \{\sum_{t=1}^T I_{jkt}(i)\left (\frac{I(A_{i,t}=j)Y_{i,t}}{\pi_j^t(Z_{i,t} )} -\theta_{jk}\right ) \right\}$.

\textit{Proof:}

We have:
\begin{align*}
    \hat{\theta}_{jk}^{\rm ipw} - \theta_{jk} & = \frac{\frac{1}{n}\sum_{i=1}^n\sum_{t=1}^T I_{jkt}(i)\cdot \frac{I(A_{i,t }=j)Y_{i,t}}{\pi_j^t(Z_{i,t})}}{\frac{1}{n}\sum_{i=1}^n\sum_{t=1}^T I_{jkt}(i)} - \theta_{jk} \\
    &= \frac{\frac{1}{n}\sum_{i=1}^n\sum_{t=1}^T I_{jkt}(i)\cdot \biggl(\frac{I(A_{i,t }=j)Y_{i,t}}{\pi_j^t(Z_{i,t})} - \theta_{jk}\biggr)}{\frac{1}{n}\sum_{i=1}^n\sum_{t=1}^T I_{jkt}(i)} \\
    &= \frac{A_n}{B_n}.
\end{align*}

From Proposition 1:
\begin{align*}
    \mathbb{E}\biggl[\sum_{t=1}^T I_{jkt}(i)\cdot \frac{I(A_{i,t }=j)Y_{i,t}}{\pi_j^t(Z_{i,t})}\biggr] = \sum_{t=1}^T \theta_{jkt}\cdot \mathbb{P}(I_{jkt}(i) = 1).
\end{align*}

and based on the definition of $\theta_{jk}$:
\begin{align*}
    \mathbb{E}\biggl[\sum_{t=1}^T I_{jkt}(i)\cdot \biggl(\frac{I(A_{i,t }=j)Y_{i,t}}{\pi_j^t(Z_{i,t})}-\theta_{jk}\biggr)\biggr] = \sum_{t=1}^T \theta_{jkt}\cdot \mathbb{P}(I_{jkt}(i) = 1) - \sum_{t=1}^T \mathbb{P}(I_{jkt}(i) = 1)\cdot \theta_{jk} = 0.
\end{align*}

Therefore, $A_n$ is an asymptotic linear estimator (ALE) with mean $0$, with influence function (IF):
\begin{align*}
    \phi_{A_n}(R_i) = \sum_{t=1}^T I_{jkt}(i)\cdot \biggl(\frac{I(A_{i,t }=j)Y_{i,t}}{\pi_j^t(Z_{i,t})} - \theta_{jk}\biggr).
\end{align*}

Moreover, $B_n$ is an ALE of $\sum_{t=1}^T\mathbb{P}(I_{jkt}(i) = 1)$, based on delta method of IF, $A_n/B_n$ is an ALE with mean 0 and IF:
\begin{align*}
    \phi^{\rm ipw}_{jk}(R_i)&=\frac{1}{\sum_{t=1}^T\mathbb{P}(I_{jkt}(i)=
     1)}\left \{ \sum_{t=1}^T I_{jkt}(i)\left(\frac{I(A_{i,t}=j)Y_{i,t}}{\pi_j^{t}(Z_{i,t})}- \theta_{jk} \right) \right \}.
\end{align*}

Therefore,  we conclude that $\hat{\theta}^{\rm ipw}_{jk}$ is an ALE of $\theta_{jk}$, with IF $\phi^{\rm ipw}_{jk}(R_i)$. Then it follows that $\sqrt{n}(\hat{\theta}_{jk}^{\text{ipw}}-\theta_{jk})\overset{d}{\to}\mathbb{N}(0,\text{Var}[\phi^{\rm ipw}_{jk}(R_i)])$. Moreover, as $n \to \infty$, $ \sqrt{n}(\hat{\pmb{\vartheta}}_{jk}^{\rm ipw} - \pmb{\vartheta}_{jk})$ converges in distribution to a bivariate normal with mean vector $\pmb{0}$ and covariance matrix ${\rm Var} ((\phi^{\rm ipw}_{jk} (R_i), \phi^{\rm ipw}_{kj}(R_i) )^T)$.

\subsection*{Proof of Theorem 1 b)}

\textbf{ Theorem 1 b):} Asymptotic distribution of the SIPW estimator $\hat{\theta}_{jk}^{\text{sipw}}$.

Under assumptions stated in Theorem 1,
\begin{align*}
    \sqrt{n}(\hat{\theta}_{jk}^{\text{sipw}}-\theta_{jk})\overset{d}{\to}\mathbb{N}(0,\text{Var}[\phi^{\rm sipw}_{jk}(R_i)]),
\end{align*}
with the IF : $\phi^{\rm sipw}_{jk}(R_i) = \frac{1}{\sum_{t=1}^T\mathbb{P}(I_{jkt}(i)=
     1)}\left \{\sum_{t=1}^T I_{jkt}(i)\left(\frac{I(A_{i,t}=j)[Y_{i,t}-\theta_{jk}]}{\pi_j^t(Z_{i,t})} \right) \right \}.$

\textit{Proof: }

We have $\hat{\theta}_{jk}^{\text{sipw}} - \theta_{jk} = \frac{\xi_{jk}}{\eta_{jk}}$, where:

\begin{align*}
    \xi_{jk}: = \frac{1}{\sum_{i=1}^n    \sum_{t = 1}^T I_{jkt}(i)}\cdot  \sum_{i=1}^n    \sum_{t = 1}^T \left\{I_{jkt}(i)\left[ \frac{I(A_{i,t} = j)(Y_{i,t}-\theta_{jk})}{\pi_{j}^t(Z_{i,t})}\right] \right\},
\end{align*}

\begin{align*}
    \eta_{jk}: = \frac{1}{\sum_{i=1}^n    \sum_{t = 1}^T I_{jkt}(i)} \cdot \sum_{i=1}^n    \sum_{t = 1}^T \left\{I_{jkt}(i)\left[ \frac{I(A_{i,t} = j)}{\pi_{j}^t(Z_{i,t})}\right] \right\}.
\end{align*}

We will first show that $\eta_{jk}\overset{p}{\to} 1$. Following the proof of Proposition 1, we have that:

\begin{align*}
    \mathbb{E}\left\{I_{jkt}(i)\left[ \frac{I(A_{i,t} = j)}{\pi_{j}^t(Z_{i,t})}\right] \right\} & = \mathbb{E}\left\{I_{jkt}(i)\cdot\mathbb{E}\left[ \frac{I(A_{i,t} = j)}{\pi_{j}^t(Z_{i,t})}\Big | \mathcal{F}_{i,t},I(T_i\geq t)\right] \right\}  = \mathbb{P}(I_{jkt}(i) = 1)
\end{align*}

Therefore, based on weak law of large number, we have $\frac{1}{n}\sum_{i=1}^n \sum_{t = 1}^T \left\{I_{jkt}(i)\left[ \frac{I(A_{i,t} = j)}{\pi_{j}^t(Z_{i,t})}\right] \right\} $ converge in probability to $ \sum_{t=1}^T \mathbb{P}(I_{jkt}(i) = 1)$, and $\frac{1}{n}\sum_{i=1}^n    \sum_{t = 1}^T I_{jkt}(i) $ converge in probability to $ \sum_{t=1}^T \mathbb{P}(I_{jkt}(i) = 1)$. Then, $\eta_{jk} $ converge in probability to $ 1$ by the continuous mapping theorem.

Following the proof of Theorem 1 a) and the derivation of $\eta_{jk}$ above:

\begin{align*}
    &\mathbb{E}\left \{\sum_{t=1}^T I_{jkt}(i)\left(\frac{I(A_{i,t}=j)(Y_{i,t}-\theta_{jk})}{\pi_j^t(Z_{i,t})} \right) \right \} \\
    =& \sum_{t=1}^T\mathbb{P}(I_{jkt}(i) = 1)\theta_{jkt} - \theta_{jk}\cdot \Big(\sum_{t=1}^T\mathbb{P}(I_{jkt}(i) = 1) \Big) = 0
\end{align*}

Therefore, $\xi_{jk}$ is asymptotically linear with mean $0$ and IF:
\begin{align*}
    \phi^{\xi}_{jk}(R_i) = \frac{1}{\sum_{t=1}^T\mathbb{P}(I_{jkt}(i)=
     1)}\left \{\sum_{t=1}^T I_{jkt}(i)\left(\frac{I(A_{i,t}=j)(Y_{i,t}-\theta_{jk})}{\pi_j^t(Z_{i,t})} \right) \right \}
\end{align*}

Since $\eta_{jk} $ converge in probability to $1$, we have $\phi^{\rm sipw}_{jk}(R_i) = \phi^{\xi}_{jk}(R_i)$ based on delta method.

\subsection*{Donsker condition in Assumption 2}

For given $j$, $k$ and $t$, there exists a class $\mathcal{F}_{jkt}$ of functions of $\pmb{X}$ such that
\begin{enumerate}[label=(\roman*),leftmargin=*, align=left]
\item $\mu_{jkt} \in \mathcal{F}_{jkt}$, where $\mu_{jkt}$ is the working model in Assumption~2,
\item $\mathbb{P}\!\left(\hat{\mu}_{jkt} \in \mathcal{F}_{jkt}\right) \to 1$ as $n \to \infty$, where $\mathbb{P}$ is the probability with respect to the randomness of $\hat{\mu}_{jkt}$ as a function of the data, and
\item
\begin{align*}
    \int_{0}^{1} \sup_{Q}\, \sqrt{ \log N\!\left(\mathcal{F}_{jkt}, \,\| \cdot \|_{L_2(Q)}, \, s \right)} \, ds < \infty,
\end{align*}

where $Q$ is any finitely supported probability distribution on the range of $\pmb{X}$, $N\!\left(\mathcal{F}_{jkt}, \| \cdot \|_{L_2(Q)}, s\right)$ is the $s$-covering number of the metric space $\big(\mathcal{F}_{jkt}, \| \cdot \|_{L_2(Q)}\big)$.
\end{enumerate}

\subsection*{Proof of Theorem 1 c)}

\textbf{Theorem 1 c): } Asymptotic distribution of the AIPW estimator $\hat{\theta}_{jk}^\text{aipw}$.

Under the assumptions stated in Theorem 1,
\begin{align*}
    \sqrt{n}[\hat{\theta}_{jk}^{\text{aipw}}-\theta_{jk}] \overset{d}{\to}\mathbb{N}(0,\text{Var}[\phi^{\rm aipw}_{jk}(R_i)])
\end{align*}

with IF:  $\phi^{\rm aipw}_{jk}(R_i) = \frac{1}{\sum_{t=1}^T\mathbb{P}(I_{jkt}(i) = 1)}\cdot  \left \{\sum_{t=1}^T I_{jkt}(i)\left (\frac{I(A_{i,t}=j)[Y_{i,t}-\mu_{jkt}(\pmb{X}_{i,t})]}{\pi_j^t(Z_{i,t} )}+ \mu_{jkt}(\pmb{X}_{i,t}) - \theta_{jk} \right ) \right\}$.

\textit{Proof:}

We first define the $\theta_{jk}^{\text{aipw}}$ as:
\begin{align*}
    \theta_{jk}^{\text{aipw}} =\frac{1}{n_{jk}} \sum_{i=1}^n \sum_{t = 1}^T \left\{  I_{jkt}(i)\left[ \frac{I(A_{i,t} = j)[Y_{i,t}-{\mu}_{jkt}(\pmb{X}_{i,t})]}{\pi_{j}^t(Z_{i,t})}+{\mu}_{jkt}(\pmb{X}_{i,t})\right] \right\},
\end{align*}

where $\mu_{jkt}(\cdot)$ is not estimated from the data. We will first show that $\sqrt{n}[{\theta}_{jk}^{\text{aipw}}-\theta_{jk}] \overset{d}{\to}\mathbb{N}(0,\text{Var}[\phi^{\rm aipw}_{jk}(R_i)])$.

Similar to the proof of Theorem 1 a):
\begin{align*}
    \theta_{jk}^{\rm aipw} - \theta_{jk} &= \frac{\frac{1}{n}\sum_{i=1}^n\sum_{t=1}^T I_{jkt}(i)\cdot \biggl(\frac{I(A_{i,t }=j)[Y_{i,t}-\mu_{jkt}(\pmb{X}_{i,t})]}{\pi_j^t(Z_{i,t})}+\mu_{jkt}(\pmb{X}_{i,t}) - \theta_{jk}\biggr)}{\frac{1}{n}\sum_{i=1}^n\sum_{t=1}^T I_{jkt}(i)} \\
    &= \frac{A^{\rm aipw}_n}{B_n},
\end{align*}
and:
\begin{align*}
    A^{\rm aipw}_n = A_n + \frac{1}{n}\sum_{i=1}^n\sum_{t=1}^T I_{jkt}(i)\cdot\Bigg(1 - \frac{I(A_{i,t} = j)}{\pi_j^t(Z_{i,t})}\Bigg)\cdot \mu_{jkt}(\pmb{X}_{i,t}).
\end{align*}

Because for all $t$, $A_{i,t}\perp \pmb{X}_{i,t}| \mathcal{F}_{i,t-1},T_i \geq t $, we have:
\begin{align*}
    &\mathbb{E}\Bigg[ I_{jkt}(i)\cdot\Bigg(1 - \frac{I(A_{i,t} = j)}{\pi_j^t(Z_{i,t})}\Bigg)\cdot \mu_{jkt}(\pmb{X}_{i,t}) \Bigg] \\
    =& \mathbb{E}\Bigg[ I_{jkt}(i)\cdot \mu_{jkt}(\pmb{X}_{i,t})\cdot \mathbb{E}\Bigg(1 - \frac{I(A_{i,t} = j)}{\pi_j^t(Z_{i,t})}\Big| \mathcal{F}_{i,t-1},\pmb{X}_{i,t},T_i \geq t\Bigg) \Bigg]\\
    =& 0,
\end{align*}
$A^{\rm aipw}_n$ is an ALE with mean $0$ and IF:
\begin{align*}
    \phi_{A^{\rm aipw}_n}(R_i) = \sum_{t=1}^T I_{jkt}(i)\cdot \biggl(\frac{I(A_{i,t }=j)[Y_{i,t}-\mu_{jkt}(\pmb{X}_{i,t})]}{\pi_j^t(Z_{i,t})}+\mu_{jkt}(\pmb{X}_{i,t}) - \theta_{jk}\biggr).
\end{align*}

Then based on delta method of IF, $A^{\rm aipw}_n/B_n$ is an ALE with mean $0$ and IF:
\begin{align*}
    \phi^{\rm aipw}_{jk}(R_i) = \frac{1}{\sum_{t=1}^T\mathbb{P}(I_{jkt}(i) = 1)}\cdot  \left \{\sum_{t=1}^T I_{jkt}(i)\left (\frac{I(A_{i,t}=j)[Y_{i,t}-\mu_{jkt}(\pmb{X}_{i,t})]}{\pi_j^t(Z_{i,t} )}+ \mu_{jkt}(\pmb{X}_{i,t}) - \theta_{jk} \right ) \right\}
\end{align*}

Therefore,  we conclude that $\theta_{jk}^{\rm aipw}$ is an ALE of $\theta_{jk}$, with IF $\phi^{\rm aipw}_{jk}(R_i)$. Then it follows that $\sqrt{n}(\theta_{jk}^{\rm aipw}-\theta_{jk})\overset{d}{\to}\mathbb{N}(0,\text{Var}[\phi^{\rm aipw}_{jk}(R_i)])$. Moreover, as $n \to \infty$, $ \sqrt{n}(\hat{\pmb{\vartheta}}_{jk}^{\rm aipw} - \pmb{\vartheta}_{jk})$ converges in distribution to a bivariate normal with mean vector $\pmb{0}$ and covariance matrix ${\rm Var} ((\phi^{\rm aipw}_{jk} (R_i), \phi^{\rm aipw}_{kj}(R_i) )^T)$.

We next show that, under the assumptions of Theorems 1 and 2, $\sqrt{n}(\hat{\theta}_{jk}^{\text{aipw}} - {\theta}_{jk}^{\text{aipw}}) = o_p(1)$.

Define $\Delta_{jk}= \hat{\theta}^{\text{aipw}}_{jk} - \theta^{\text{aipw}}_{jk}$ and define  $\widetilde{\Delta}_{jk}$ as follows:
\begin{align*}
    \Delta_{jk}  &=\frac{1}{\frac{1}{n}\sum_{i=1}^n E_{jk}^{(i)}} \cdot \frac{1}{n}\sum_{i=1}^n \left\{ \sum_{t=1}^T I_{jkt}(i)\left [\left (1-\frac{I(A_{i,t}=j)}{\pi_j^{t}(Z_{i,t})} \right )\cdot[\hat{\mu}_{jk}(\pmb{X}_{i,t})- {\mu}_{jk}(\pmb{X}_{i,t})]\right ]\right\},
\end{align*}

\begin{align*}
    \widetilde{\Delta}_{jk}=\frac{1}{\sum_{t=1}^T\mathbb{P}(I_{jkt}(i)=1)} \cdot \frac{1}{n}\sum_{i=1}^n \left\{ \sum_{t=1}^T I_{jkt}(i)\left [\left (1-\frac{I(A_{i,t}=j)}{\pi_j^{t}(Z_{i,t})} \right )\cdot[\hat{\mu}_{jk}(\pmb{X}_{i,t})- {\mu}_{jk}(\pmb{X}_{i,t})]\right ]\right\}.
\end{align*}

By the Cauchy–Schwarz inequality and Assumption 2, we obtain:
\begin{align*}
    &\Bigg |\frac{1}{n}\sum_{i=1}^n\left\{ \sum_{t=1}^T I_{jkt}(i)\left [\left (1-\frac{I(A_{i,t}=j)}{\pi_j^{t}(Z_{{i,t}})} \right )\cdot[\hat{\mu}_{jkt}(\pmb{X}_{i,t})- {\mu}_{jkt}(\pmb{X}_{i,t})]\right ]\right\}\Bigg | \\
    \leq& \left\{\frac{1}{n} \sum_{i=1}^n \sum_{t=1}^T  I_{jkt}(i) \cdot \left (1-\frac{I(A_{i,t}=j)}{\pi_j^{t}(Z_{i,t})} \right )^2 \right   \}^{1/2}\cdot \left\{\frac{1}{n} \sum_{i=1}^n \sum_{t=1}^T [\hat{\mu}_{jkt}(\pmb{X}_{i,t})- {\mu}_{jkt}(\pmb{X}_{i,t})]^2\right\}^{1/2} \\
    =& \sqrt{O_p(1)}\cdot \sqrt{o_p(1)} = o_p(1).
\end{align*}

Moreover, we have:
\begin{align*}
    \sqrt{n}\left (\frac{1}{\frac{1}{n}\sum_{i=1}^n E_{jk}^{(i)}} - \frac{1}{\sum_{t=1}^T\mathbb{P}(I_{jkt}(i)=1)} \right )
    =\frac{\sqrt{n}[\sum_{t=1}^T\mathbb{P}(I_{jkt}(i)=1) - \frac{1}{n}\sum_{i=1}^n E_{jk}^{(i)}]}{\{\frac{1}{n}\sum_{i=1}^n E_{jk}^{(i)}\}\cdot \{\sum_{t=1}^T\mathbb{P}(I_{jkt}(i)=1)\}} = O_p(1)
\end{align*}

Putting the two displays above together, we obtain: $\sqrt{n}(\Delta_{jk} - \widetilde{\Delta}_{jk}) = o_p(1) \cdot O_p(1) = o_p(1).$

Next, define:
\begin{align*}
    g_n:= \frac{\sum_{t=1}^T I_{jkt}(i)\left [\left (1-\frac{I(A_{i,t}=j)}{\pi_j^{t}(i)} \right )\cdot[\hat{\mu}_{jkt}(\pmb{X}_{i,t})- {\mu}_{jkt}(\pmb{X}_{i,t})]\right ]}{\sum_{t=1}^T\mathbb{P}(I_{jkt}(i)=1)}
\end{align*}

Again by the Cauchy–Schwarz inequality and Assumption 2, we have:
\begin{align*}
    \mathbb{E} (g_n^2) &\leq \frac{\mathbb{E}\Bigg\{ \Bigg(\sum_{t=1}^T  I_{jkt}(i) \cdot \Big(1-\frac{I(A_{i,t}=j)}{\pi_j^{t}(Z_{i,t})} \Big )^2 \Bigg) \cdot \Big( \sum_{t=1}^T [\hat{\mu}_{jkt}(\pmb{X}_{i,t})- {\mu}_{jkt}(\pmb{X}_{i,t})]^2\Big)\Bigg\}}{[\sum_{t=1}^T\mathbb{P}(I_{jkt}(i)=1)]^2} \\
    &\leq \frac{\mathbb{E}\Bigg\{ \Bigg(\sum_{t=1}^T  I_{jkt}(i) \cdot \Big(1-\frac{I(A_{i,t}=j)}{\pi_j^{t}(Z_{i,t})} \Big )^2 \Bigg)^2\Bigg\}^{1/2} \cdot \mathbb{E}\Bigg\{\Big( \sum_{t=1}^T [\hat{\mu}_{jkt}(\pmb{X}_{i,t})- {\mu}_{jkt}(\pmb{X}_{i,t})]^2\Big)^2\Bigg\}^{1/2}}{[\sum_{t=1}^T\mathbb{P}(I_{jkt}(i)=1)]^2}\\
    &= o_p(1) 
\end{align*}

Therefore $\mathbb{E}(g_n^2) $ converge in probability to $ 0$. If $\hat{\mu}_{jk}$ is obtained from a finite-dimensional parametric model or satisfies the Donsker condition, then  $g_n$ belongs to a $P_0$-Donsker class,. Hence, by Lemma 19.24 in van der Vaart (1998), we have $\sqrt{n}\tilde{\Delta}_{jk} = o_p(1)$. Consequently, $\sqrt{n}{\Delta}_{jk} = o_p(1)$ and $\sqrt{n}[\hat{\theta}_{jk}^{\text{aipw}}-\theta_{jk}] $ converge in distribution to $\mathbb{N}(0,\text{Var}[\phi^{\rm aipw}_{jk}(R_i)])$.

\subsection*{Proof of Theorem 1 d)}

\textbf{Theorem 1 d): } Asymptotic distribution of  the PS estimator $\hat{\theta}_{jk}^{\text{ps}}$.

Under the assumptions of Theorem 1,
\begin{align*}
    \sqrt{n}(\hat{\theta}_{jk}^{\text{ps}} - \theta_{jk})\overset{d}{\to}\mathbb{N}(0,\text{Var}[\phi^{\rm ps}_{jk}(R_i)])
\end{align*}

with IF: $\phi^{\rm ps}_{jk}(R_i) = \frac{1}{\sum_{t=1}^T\mathbb{P}(I_{jkt}(i) = 1)}\cdot  \sum_{t=1}^T\Big \{ \sum_{h=1}^{H_{jkt}}\Big[\frac{I(A_{i,t} = j,G_{i,t} = h)}{\mathbb{P}(A_{i,t} = j|G_{i,t}=h)}[Y_{i,t} - \mathbb{E}(Y_{i,t}|A_{i,t} = j,G_{i,t}=h)]+I(G_{i,t}=h)[\mathbb{E}(Y_{i,t}|A_{i,t} = j,G_{i,t}=h)-\theta_{jk}] \Big]\Big\}.$

\textit{Proof:} 

Since $n(\mathcal{I}_{jkt}^{(h)}) = \sum_{i=1}^n I(G_{i,t} = h) $ and $n_j(\mathcal{I}_{jkt}^{(h)}) = \sum_{i=1}^n I(A_{i,t} = j,G_{i,t} = h)$, we can decompose $\hat{\theta}_{jk}^{\text{ps}} - \theta_{jk} = U_{jk}+V_{jk}$, where $U_{jk}$ and $V_{jk}$ are defined as follows:
\begin{align*}
    U_{jk} &=  \frac{1}{n_{jk}}\sum_{t=1}^T\Bigg \{\sum_{h=1}^{H_{jkt}}\frac{\sum_{i=1}^n I(G_{i,t} = h)}{\sum_{i=1}^nI(A_{i,t} = j,G_{i,t} = h)}\\
    &\qquad \times \{\sum_{i=1}^nI(A_{i,t} = j,G_{i,t} = h)\cdot [Y_{i,t} - \mathbb{E}(Y_{i,t}|A_{i,t} = j,G_{i,t} = h)]\}  \Bigg\},
\end{align*}

\begin{align*}
    V_{jk} &=  \frac{1}{n_{jk}}\sum_{t=1}^T\Bigg \{\sum_{h=1}^{H_{jkt}}\frac{\sum_{i=1}^n I(G_{i,t} = h)}{\sum_{i=1}^nI(A_{i,t} = j,G_{i,t} = h)}\\
    &\qquad\times \{\sum_{i=1}^nI(A_{i,t} = j,G_{i,t} = h)\cdot [\mathbb{E}(Y_{i,t}|A_{i,t} = j,G_{i,t} = h)-\theta_{jk}]\}  \Bigg\}.
\end{align*}

First, note that $V_{jk}$ can be rewritten as:
\begin{align*}
     V_{jk} &= \frac{1}{n_{jk}}\sum_{t=1}^T\Bigg \{\sum_{h=1}^{H_{jkt}}\frac{\sum_{i=1}^n I(G_{i,t} = h)}{\sum_{i=1}^nI(A_{i,t} = j,G_{i,t} = h)} \\
     &\qquad\times\{\sum_{i=1}^nI(A_{i,t} = j,G_{i,t} = h)\}\cdot [\mathbb{E}(Y_{i,t}|A_{i,t} = j,G_{i,t} = h)-\theta_{jk}]  \Bigg\}\\
     &=\frac{1}{n_{jk}}\sum_{t=1}^T \left\{\sum_{h=1}^{H_{jkt}}\Bigg[[\sum_{i=1}^n I(G_{i,t} = h)]\cdot [\mathbb{E}(Y_{i,t}|A_{i,t} = j,G_{i,t} = h)-\theta_{jk}]\Bigg]\right \} \\
     &= \frac{1}{n_{jk}}\sum_{t=1}^T \left\{ \sum_{i=1}^n\sum_{h=1}^{H_{jkt}} I(G_{i,t}=h)\cdot [\mathbb{E}(Y_{i,t}|A_{i,t} = j,G_{i,t} = h)-\theta_{jk}]\right\} \\
     & = \frac{1}{\sum_{i=1}^n \sum_{t=1}^T I_{jkt}(i)}\cdot \sum_{i=1}^n\left\{ \sum_{t=1}^T \sum_{h=1}^{H_{jkt}} I(G_{i,t}=h)\cdot [\mathbb{E}(Y_{i,t}|A_{i,t} = j,G_{i,t} = h)-\theta_{jk}]\right\}.
\end{align*}

Moreover, observe that the expectation of the bracketed term satisfies:
\begin{align*}
    &\mathbb{E}\left\{  \sum_{t=1}^T \sum_{h=1}^{H_{jkt}} I(G_{i,t}=h)\cdot [\mathbb{E}(Y_{i,t}|A_{i,t} = j,G_{i,t} = h)-\theta_{jk}]\right\} \\
    =& \sum_{t=1}^T\mathbb{E}\left\{I_{jkt}(i)\cdot  \sum_{h=1}^{H_{jkt}} \Bigg[I(G_{i,t}=h)\cdot [\mathbb{E}(Y_{i,t}|A_{i,t} = j,G_{i,t} = h)-\theta_{jk}]\Bigg]\right\} \\
    =&\sum_{t=1}^T\mathbb{E}\left\{I_{jkt}(i)\cdot  \sum_{h=1}^{H_{jkt}} I(G_{i,t}=h)\cdot \mathbb{E}(Y_{i,t}|A_{i,t} = j,G_{i,t} = h)\right\}- [\sum_{t=1}^T \mathbb{P}(I_{jkt}(i)=1)]\cdot \theta_{jk}.
\end{align*}

Consider the first term in the display above:
\begin{align*}
    &\sum_{t=1}^T\mathbb{E}\left\{I_{jkt}(i)\cdot  \sum_{h=1}^{H_{jkt}} \Bigg[I(G_{i,t}=h)\cdot \mathbb{E}(Y_{i,t}|A_{i,t} = j,G_{i,t} = h)\Bigg]\right\} \\
    =&\sum_{t=1}^T\mathbb{E}\left\{\sum_{h=1}^{H_{jkt}} \Bigg[I(G_{i,t}=h)\cdot \mathbb{E}(Y_{i,t}|A_{i,t} = j,G_{i,t} = h)\Bigg]\Big | I_{jkt}(i) = 1\right\}\times \mathbb{P}(I_{jkt}(i) = 1)\\
    =& \sum_{t=1}^T\mathbb{E}\Big\{ \mathbb{E}(Y_{i,t}|A_{i,t} = j,G_{i,t} )\Big | I_{jkt}(i) = 1\Big\}\times \mathbb{P}(I_{jkt}(i) = 1) \\
    =& \sum_{t=1}^T\mathbb{E}\Big\{ \mathbb{E}(Y_{i,t}^{(\bar{\pmb{A}}_{i,t-1},j ) }|G_{i,t} )\Big | I_{jkt}(i) = 1\Big\}\times \mathbb{P}(I_{jkt}(i) = 1) \\
    =& \sum_{t=1}^T\mathbb{E}\Big\{ Y_{i,t}^{(\bar{\pmb{A}}_{i,t-1},j ) } \Big | I_{jkt}(i) = 1\Big\}\times \mathbb{P}(I_{jkt}(i) = 1) \\
    =& \sum_{t=1}^T \theta_{jkt} \times \mathbb{P}(I_{jkt}(i) = 1)\\
    =& \;\theta_{jk}\cdot \Big (\sum_{t=1}^T \mathbb{P}(I_{jkt}(i )=1)\Big ).
\end{align*}

Therefore, $\mathbb{E}\left\{  \sum_{t=1}^T \sum_{h=1}^{H_{jkt}} I(G_{i,t}=h)\cdot \allowbreak [\mathbb{E}(Y_{i,t}|A_{i,t} = j,G_{i,t} = h)-\theta_{jk}]\right\} = 0$.

Based on the delta method on IF, $V_{jk}$ is asymptotic linear with mean $0$ and IF:
\begin{align*}
    \phi_v (R_i): = \frac{1}{\sum_{t=1}^T \mathbb{P}(I_{jkt}(i)=1)}\cdot \sum_{t=1}^T \Big \{\sum_{h=1}^{H_{jkt}} I(G_{i,t} = h)\cdot[\mathbb{E}(Y_{i,t}|A_{i,t} = j,G_{i,t} = h)-\theta_{jk}]\Big \}
\end{align*}

Next, for each stratum $h$ and episode $t$, define:
\begin{align*}
    u_{h,t}&: = \frac{\sum_{i=1}^nI(A_{i,t} = j,G_{i,t} = h)\cdot\{ Y_{i,t} - \mathbb{E}(Y_{i,t}|A_{i,t} = j,G_{i,t} = h)\}}{\sum_{i=1}^nI(A_{i,t} = j,G_{i,t} = h)} 
\end{align*}

To verify that the numerator is mean zero, note that:
\begin{align*}
    &\mathbb{E}\Big\{I(A_{i,t} = j,G_{i,t} = h)\cdot\{ Y_{i,t} - \mathbb{E}(Y_{i,t}|A_{i,t} = j,G_{i,t} = h)\}\Big\} \\
    =& \mathbb{E}\Bigg\{\mathbb{E}\Big\{I(A_{i,t} = j,G_{i,t} = h)\cdot\{ Y_{i,t} - \mathbb{E}(Y_{i,t}|A_{i,t} = j,G_{i,t} = h)\}\Big| A_{i,t},G_{i,t}\Big\}\Bigg\}\\
    =& \mathbb{E}\Big \{I(A_{i,t} = j,G_{i,t} = h)\cdot \mathbb{E}\{ Y_{i,t} - \mathbb{E}(Y_{i,t}|A_{i,t} = j,G_{i,t} = h)\Big | A_{i,t} = j,G_{i,t} = h\} \Big\} \\
    =& 0.
\end{align*}

Therefore, for each $ h\text{ and }t, u_{h,t}$ is asymptotic linear with mean $0$ and IF:
\begin{align*}
    \phi_{h,t}(R_i):= \frac{I(A_{i,t} = j,G_{i,t} = h)}{\mathbb{P}(A_{i,t}=j,G_{i,t}=h)}\cdot [Y_{i,t} - \mathbb{E}(Y_{i,t}|A_{i,t} = j,G_{i,t} = h)],
\end{align*}

and $\sqrt{n}u_{h,t} = \frac{1}{\sqrt{n}}\sum_{i=1}^n\phi_{h,t}(R_i) +o_p(1)$.

For notation, define:
\begin{align*}
    \hat{\omega}(h,t):= \frac{\sum_{i=1}^nI(G_{i,t}=h)}{\sum_{i=1}^n\sum_{t=1}^TI_{jkt}(i)}\text{ and }\omega(h,t) = \frac{\mathbb{P}(G_{i,t}=h)}{\sum_{t=1}^T\mathbb{P}(I_{jkt}(i)=1)}.
\end{align*}

With these definitions, we can write:
\begin{align*}
    \sqrt{n}U_{jk} &= \sqrt{n}\sum_{t=1}^T \sum_{h=1}^{H_{jkt}}\hat{\omega}(h,t)u_{h,t} \\
    &=\sum_{t=1}^T \sum_{h=1}^{H_{jkt}}[\hat{\omega}(h,t)- \omega(h,t)]\sqrt{n}u_{h,t} + \sum_{t=1}^T \sum_{h=1}^{H_{jkt}} w(h,t)\cdot \sqrt{n}u_{h,t}
\end{align*}

Since $\hat{\omega}(h,t)- \omega(h,t) =o_P(1),\sqrt{n}u_{h,t} = O_p(1)\text{ and }T\text{ and }H_{jkt} \text{ are finite for all }t$, it follows that $U_{jk}$ admits the representation:
\begin{align*}
     \sqrt{n}U_{jk} &= \sum_{t=1}^T \sum_{h=1}^{H_{jkt}} w(h,t)\cdot \sqrt{n}u_{h,t} +o_p (1) \\
     &=\frac{1}{\sqrt{n}}\sum_{i=1}^n\left\{\sum_{t=1}^T \sum_{h=1}^{H_{jkt}}w(h,t)\phi_{h,t}(R_i) \right\}+o_p(1) \\
     &= \frac{1}{\sqrt{n}}\sum_{i=1}^n\Bigg\{\sum_{t=1}^T \sum_{h=1}^{H_{jkt}}\frac{\mathbb{P}(G_{i,t}=h)}{\sum_{t=1}^T\mathbb{P}(I_{jkt}(i)=1)}  \\
     & \;\;\;\;\;\;\;\;\;\;\;\;\;\;\;\;\;\;\;\;\;\;\;\times \frac{I(A_{i,t} = j,G_{i,t} = h)}{\mathbb{P}(A_{i,t}=j,G_{i,t}=h)}\cdot [Y_{i,t} - \mathbb{E}(Y_{i,t}|A_{i,t} = j,G_{i,t}=h)] \Bigg\}+o_p(1)\\
     &= \frac{1}{\sqrt{n}}\sum_{i=1}^n\Bigg\{\sum_{t=1}^T \sum_{h=1}^{H_{jkt}}\frac{1}{\sum_{t=1}^T\mathbb{P}(I_{jkt}(i)=1)}  \\
     & \;\;\;\;\;\;\;\;\;\;\;\;\;\;\;\;\;\;\;\;\;\;\;\Bigg.\times \frac{I(A_{i,t} = j,G_{i,t} = h)}{\mathbb{P}(A_{i,t}=j|G_{i,t}=h)}\cdot [Y_{i,t} - \mathbb{E}(Y_{i,t}|A_{i,t} = j,G_{i,t}=h)] \Bigg\}+o_p(1)\\
\end{align*}

Therefore, $U_{jk}$ is aysmptotic linear with mean 0 and  IF:
\begin{align*}
    \phi_u(R_i) = \frac{1}{\sum_{t=1}^T\mathbb{P}(I_{jkt}(i)=1)} \sum_{t=1}^T\left\{\sum_{h=1}^{H_{jkt}}\Bigg[ \frac{I(A_{i,t} = j,G_{i,t} = h)}{\mathbb{P}(A_{i,t}=j|G_{i,t}=h)}\cdot [Y_{i,t} - \mathbb{E}(Y_{i,t}|A_{i,t} = j,G_{i,t}=h)]\Bigg] \right\}
\end{align*}

Combining the linear representations for parts $U_{jk}$ and $V_{jk}$, we conclude that $\hat{\theta}_{jk}^{\text{ps}}$ is an ALE of $\theta_{jk}$, with IF:
\begin{align*}
    \phi^{\rm ps}_{jk}(R_i) &= \phi_u(R_i) + \phi_v(R_i)\\
    &= \frac{1}{\sum_{t=1}^T\mathbb{P}(I_{jkt}(i)=1)} \sum_{t=1}^T\Bigg\{\sum_{h=1}^{H_{jkt}} \frac{I(A_{i,t} = j,G_{i,t} = h)}{\mathbb{P}(A_{i,t}=j|G_{i,t}=h)}\cdot [Y_{i,t} - \mathbb{E}(Y_{i,t}|A_{i,t} = j,G_{i,t}=h)] \\
    &  +I(G_{i,t} = h)\cdot [\mathbb{E}(Y_{i,t}|A_{i,t} = j,G_{i,t}=h) - \theta_{jk}]\Bigg\}
\end{align*}

\subsection*{Proof of Theorem 1 e)}

\textbf{Theorem 1 e)} Asymptotic distribution of the APS estimator $\hat{\theta}_{jk}^{\text{aps}}$.

Under the assumptions of Theorem 1,
\begin{align*}
    \sqrt{n}(\hat{\theta}_{jk}^{\text{aps}} - \theta_{jk}) \overset{d}{\to} \mathbb{N}(0,\mathrm{Var}[\phi^{\rm aps}_{jk}(R_i)])
\end{align*}

with IF: $\phi^{\rm aps}_{jk}(R_i)=\frac{1}{\sum_{t=1}^T\mathbb{P}(I_{jkt}(i)=1)}\times \sum_{t=1}^T\Bigg\{\sum_{h=1}^{H_{jkt}} \frac{I(A_{i,t} = j,G_{i,t} = h)}{\mathbb{P}(A_{i,t}=j|G_{i,t}=h)}\cdot \Big\{Y_{i,t} -\mu_{jkt}(\pmb{X}_{i,t})+\bar{\mu}_{jkt}(h) - \mathbb{E}(Y_{i,t}|A_{i,t} = j,G_{i,t}=h)\Big\} +I(G_{i,t} = h)\cdot\Big[\mathbb{E}(Y_{i,t}|A_{i,t}=j,G_{i,t} = h) +\mu_{jkt}(\pmb{X}_{i,t}) - \bar{\mu}_{jkt}(h)-\theta_{jk}\Big]\Bigg\}$. 

\textit{Proof:}

First define: 
\begin{align*}
    {\theta}^{\text{APS}}_{jk} &=  \frac{1}{n_{jk}}\sum_{t=1}^T\left \{\sum_{h=1}^{H_{jkt}}\frac{\sum_{i=1}^n I(G_{i,t} = h)}{\sum_{i=1}^nI(A_{i,t} = j,G_{i,t} = h)}\{\sum_{i=1}^nI(A_{i,t} = j,G_{i,t} = h)\cdot [Y_{i,t}-{\mu}_{jkt}(\pmb{X}_{i,t})]\} \right\} \\ &+ \frac{1}{n_{jk}}\sum_{i=1}^n\sum_{t=1}^T I_{jkt}(i)\cdot {\mu}_{jkt}(\pmb{X}_{i,t}),
\end{align*}
where $\mu_{jkt}(\cdot)$ is treated as fixed and is not estimated from the data. As in the proof of Theorem 1 d), we decompose $\theta_{jk}^{\text{aps}} - \theta_{jk} = U_{jk}^a+V_{jk}^a$, where:
\begin{align*}
    U_{jk}^a =  \frac{1}{n_{jk}}&\sum_{t=1}^T\left \{  \sum_{h=1}^{H_{jkt}}\frac{\sum_{i=1}^n I(G_{i,t} = h)}{\sum_{i=1}^nI(A_{i,t} = j,G_{i,t} = h)}\right. \\
    &\left. \times\Big\{\sum_{i=1}^nI(A_{i,t} = j,G_{i,t} = h)\cdot [Y_{i,t} -\mu_{jkt}(\pmb{X}_{i,t})+\bar{\mu}_{jkt}(h) - \mathbb{E}(Y_{i,t}|A_{i,t} = j,G_{i,t} = h)]\Big\}  \right\},
\end{align*}

where $\bar{\mu}_{jkt}(h):= \mathbb{E}[\mu_{jkt}(\pmb{X}_{i,t})|G_{i,t} = h]$, and define:
\begin{align*}
    V_{jk}^a:= \frac{1}{n_{jk}}&\sum_{t=1}^T\left \{  \sum_{h=1}^{H_{jkt}}\frac{\sum_{i=1}^n I(G_{i,t} = h)}{\sum_{i=1}^nI(A_{i,t} = j,G_{i,t} = h)} \right. \\
    &\times \left. \Big\{\sum_{i=1}^nI(A_{i,t} = j,G_{i,t} = h)\cdot [ \mathbb{E}(Y_{i,t}|A_{i,t} = j,G_{i,t} =h) - \bar{\mu}_{jkt}(h) - \theta_{jk}]\Big\}  \right\}  \\
    +& \frac{1}{|\mathcal{I}_{jk}|}\sum_{i=1}^n\sum_{t=1}^T I_{jkt}(i)\cdot {\mu}_{jkt}(\pmb{X}_{i,t}).
\end{align*}

We first note that:
\begin{align*}
    V_{jk}^a = V_{jk}+ \frac{1}{|\mathcal{I}_{jk}|}\sum_{i=1}^n\sum_{t=1}^T \sum_{h=1}^{H_{jkt}}I(G_{i,t} = h)\cdot[\mu_{jkt}(\pmb{X}_{i,t}) - \bar{\mu}_{jkt}(h)].
\end{align*}

Since $\mathbb{E}\{I(G_{i,t} = h)\cdot[\mu_{jkt}(\pmb{X}_{i,t}) - \bar{\mu}_{jkt}(h)]\} = 0$ for all episode $t$ and stratum $h$, we have that $V_{jk}^a$ is asymptotically linear with mean $0$ and IF:
\begin{align*}
    \phi_v^a(R_i) := \frac{1}{\sum_{t=1}^T \mathbb{P}(I_{jkt}(i)=1)}\cdot \sum_{t=1}^T \Big \{\sum_{h=1}^{H_{jkt}} I(G_{i,t} = h)\cdot[&\mathbb{E}(Y_{i,t}|A_{i,t} = j,G_{i,t} = h) +\mu_{jkt}(\pmb{X}_{i,t}) \\
    &- \bar{\mu}_{jkt}(h)-\theta_{jk}]\Big \}
\end{align*}

Next, for each stratum $h$ and  episode $t$, define:
\begin{align*}
    u_{h,t}^a &:= \frac{\sum_{i=1}^nI(A_{i,t} = j,G_{i,t} = h)\cdot \{Y_{i,t} -\mu_{jkt}(\pmb{X}_{i,t})+\bar{\mu}_{jkt}(h) - \mathbb{E}(Y_{i,t}|A_{i,t} = j,G_{i,t} = h)\}}{\sum_{i=1}^nI(A_{i,t} = j,G_{i,t} = h)} \\
\end{align*}

By Assumption 1 and the balancing property of the propensity score, $A_{i,t}\perp \pmb{X}_{i,t}|\pi_j^t(Z_{i,t})$, and hence $A_{i,t}\perp \pmb{X}_{i,t}|\pi_j^t(Z_{i,t}),G_{i,t}$. Moreover, because $\pi_j^t(Z_{i,t})$ is constant within each stratum indexed by $G_{i,t}$, conditioning on $G_{i,t}$ is sufficient, so $A_{i,t}\perp \pmb{X}_{i,t} | G_{i,t}$.

Taking expectations of the numerator yields:
\begin{align*}
    &\mathbb{E}\Big\{I(A_{i,t} = j,G_{i,t} = h)\cdot \{Y_{i,t} -\mu_{jkt}(\pmb{X}_{i,t})+\bar{\mu}_{jkt}(h) - \mathbb{E}(Y_{i,t}|A_{i,t} = j,G_{i,t}=h)\}\Big\}\\
    =& \mathbb{E}\Big\{I(A_{i,t} = j,G_{i,t} = h)\cdot \mathbb{E}\{Y_{i,t} -\mu_{jkt}(\pmb{X}_{i,t})+\bar{\mu}_{jkt}(h) - \mathbb{E}(Y_{i,t}|A_{i,t} = j,G_{i,t} = h)\Big |A_{i,t}=j,G_{i,t}=h\}\Big\}\\
    =& \mathbb{E}\{I(A_{i,t} = j,G_{i,t} = h)\cdot \mathbb{E}\{ -\mu_{jkt}(\pmb{X}_{i,t})+\bar{\mu}_{jkt}(h) \Big |A_{i,t}=j,G_{i,t}=h\}\} =0
\end{align*}

The third equality follows from $A_{i,t}\perp \pmb{X}_{i,t} | G_{i,t}$.

Therefore, for each $h\text{ and }t,\;u_{h,t}^a$ is an asymptotically linear with mean $0$ and IF:
\begin{align*}
    \phi_{h,t}^a(R_i):= \frac{I(A_{i,t} = j,G_{i,t} = h)}{\mathbb{P}(A_{i,t}=j,G_{i,t}=h)} \cdot \{Y_{i,t} -\mu_{jkt}(\pmb{X}_{i,t})+\bar{\mu}_{jkt}(h) - \mathbb{E}(Y_{i,t}|A_{i,t} = j,G_{i,t}=h)\}.
\end{align*}

Moreover, $\sqrt{n}u_{h,t}^a = \frac{1}{\sqrt{n}}\sum_{i=1}^n\phi_{h,t}^a(R_i) + o_p(1)$.

Following the same arguments as in the proof of Theorem 1 d), we obtain:
\begin{align*}
    \sqrt{n}U_{jk}^a &= \sum_{t=1}^T \sum_{h=1}^{H_{jkt}} w(h,t)\cdot \sqrt{n}u_{h,t}^a +o_p (1) \\
    &=\frac{1}{\sqrt{n}}\sum_{i=1}^n\left\{\sum_{t=1}^T \sum_{h=1}^{H_{jkt}} \omega(h,t)\phi_{h,t}^a(R_i)\right\}+o_p(1) \\
    &= \frac{1}{\sqrt{n}}\sum_{i=1}^n\Bigg\{\sum_{t=1}^T \sum_{h=1}^{H_{jkt}}\frac{1}{\sum_{t=1}^T\mathbb{P}(I_{jkt}(i)=1)} \cdot \frac{I(A_{i,t} = j,G_{i,t} = h)}{\mathbb{P}(A_{i,t}=j|G_{i,t}=h)}  \\
     & \;\;\;\;\;\;\;\;\;\;\;\;\;\;\;\;\;\;\;\;\;\times  \{Y_{i,t} -\mu_{jkt}(\pmb{X}_{i,t})+\bar{\mu}_{jkt}(h) - \mathbb{E}(Y_{i,t}|A_{i,t} = j,G_{i,t}=h)\} \Bigg\}+o_p(1)
\end{align*}

Hence, $U_{jk}^a$ is asymptotically linear with mean $0$ and IF:
\begin{align*}
    \phi_u^a(R_i) :=  \frac{1}{\sum_{t=1}^T\mathbb{P}(I_{jkt}(i)=1)}  \sum_{t=1}^T\Bigg \{&\sum_{h=1}^{H_{jkt}} \frac{I(A_{i,t} = j,G_{i,t} = h)}{\mathbb{P}(A_{i,t}=j|G_{i,t}=h)}  \\
    &\times \{Y_{i,t} -\mu_{jkt}(\pmb{X}_{i,t})+\bar{\mu}_{jkt}(h) - \mathbb{E}(Y_{i,t}|A_{i,t} = j,G_{i,t}=h)\}\Bigg \}
\end{align*}

Finally, combining the expansions for $U_{jk}^a$ and $V_{jk}^a$ yields the IF of $\theta^{\text{aps}}_{jk}$ as $\phi_u^a(R_i) + \phi^a_v(R_i):$
\begin{align*}
    \phi^{\rm aps}_{jk}(R_i)
    = &\frac{1}{\sum_{t=1}^T\mathbb{P}(I_{jkt}(i)=1)}\times \\
    \sum_{t=1}^T\Big\{&\sum_{h=1}^{H_{jkt}} \frac{I(A_{i,t} = j,G_{i,t} = h)}{\mathbb{P}(A_{i,t}=j|G_{i,t}=h)}\cdot \{Y_{i,t} -\mu_{jkt}(\pmb{X}_{i,t})+\bar{\mu}_{jkt}(h) - \mathbb{E}(Y_{i,t}|A_{i,t} = j,G_{i,t}=h)\} \\
    &  +I(G_{i,t} = h)\cdot[\mathbb{E}(Y_{i,t}|A_{i,t} = j,G_{i,t} = h) +\mu_{jkt}(\pmb{X}_{i,t}) - \bar{\mu}_{jkt}(h)-\theta_{jk}]\Big\}
\end{align*}

Next, we show that $\sqrt{n}(\hat{\theta}_{jk}^{\text{aps}} - {\theta}_{jk}^{\text{aps}}) = o_p(1)$. Define $\Delta_{jk}^a = \hat{\theta}_{jk}^{\text{aps}} - {\theta}_{jk}^{\text{aps}}$ and $\tilde{\Delta}_{jk}^a$ as follows:
\begin{align*}
    \Delta_{jk}^a  = \frac{1}{n_{jk}}\sum_{i=1}^n\left\{ \sum_{t=1}^T I_{jkt}(i)\left[ \left(1 -\sum_{h=1}^{H_{jkt}} \frac{I(G_{i,t} = h)\cdot I(A_{i,t} = j)}{\hat{\mathbb{P}}(A_{i,t}=j|G_{i,t} = h)}\right)\cdot[\hat{\mu}_{jkt}(\pmb{X}_{i,t}) -{\mu}_{jkt}(\pmb{X}_{i,t})]\right]\right\},
\end{align*}
where $\hat{\mathbb{P}}(A_{i,t}=j|G_{i,t} = h) = [\sum_{i=1}^nI(G_{i,t} = h)\cdot I(A_{i,t} = j)]/[\sum_{i=1}^nI(G_{i,t} = h)]$; and define:
\begin{align*}
    \tilde{\Delta}_{jk}^a  &= \frac{1}{\sum_{t=1}^T\mathbb{P}(I_{jkt}(i)=1)} \\
    &\times\frac{1}{n}\sum_{i=1}^n\left\{ \sum_{t=1}^T I_{jkt}(i)\left[ \left(1 -\sum_{h=1}^{H_{jkt}} \frac{I(G_{i,t} = h)\cdot I(A_{i,t} = j)}{\hat{\mathbb{P}}(A_{i,t}=j|G_{i,t} = h)}\right)\cdot[\hat{\mu}_{jkt}(\pmb{X}_{i,t}) -{\mu}_{jkt}(\pmb{X}_{i,t})]\right]\right\},
\end{align*}

By the Cauchy–Schwarz inequality, we obtain:
\begin{align*}
    &\Bigg |\frac{1}{n}\sum_{i=1}^n\left\{ \sum_{t=1}^T I_{jkt}(i)\left[ \left(1 -\sum_{h=1}^{H_{jkt}} \frac{I(G_{i,t} = h,A_{i,t} = j)}{\hat{\mathbb{P}}(A_{i,t}=j|G_{i,t} = h)}\right)\cdot[\hat{\mu}_{jk}(\pmb{X}_{i,t}) -{\mu}_{jk}(\pmb{X}_{i,t})]\right]\right\}\Bigg | \\
    \leq &\left \{ \frac{1}{n}\sum_{i=1}^n \sum_{t=1}^T I_{jkt}(i)\cdot\left(1 -\sum_{h=1}^{H_{jkt}} \frac{I(G_{i,t} = h,A_{i,t} = j)}{\hat{\mathbb{P}}(A_{i,t}=j|G_{i,t} = h)}\right)^2 \right \}^{1/2} \\
    & \times \left \{ \frac{1}{n}\sum_{i=1}^n \sum_{t=1}^T [\hat{\mu}_{jkt}(\pmb{X}_{i,t}) -{\mu}_{jkt}(\pmb{X}_{i,t})]^2 \right \}^{1/2} \\
    =& \sqrt{O_p(1)}\cdot \sqrt{o_p(1)} = o_p(1);
\end{align*}

Hence, the displayed term is $o_p(1)$. Moreover, in the proof of Theorem 1 c) we showed that $\sqrt{n}(\frac{1}{n_{jk}/n} - \frac{1}{\sum_{t=1}^T\mathbb{P}(I_{jkt}(i)=1)}) = O_p(1)$. Putting these pieces together, we have $\sqrt{n}(\Delta_{jk}^a - \tilde{\Delta}_{jk}^a) = o_p(1)$. As in the proof of Theorem 1 c), we apply Lemma 19.24 of van der Vaart (1998) to conclude that $\sqrt{n}\tilde{\Delta}_{jk}^a = o_p(1)$. Therefore, $\sqrt{n}{\Delta}_{jk}^a = o_p(1)$. Consequently, $\sqrt{n}[\hat{\theta}_{jk}^{\text{aps}}-\theta_{jk}]\overset{d}{\to}\mathbb{N}(0,\mathrm{Var}[\phi^{\rm aps}_{jk}(R_i)])$.

\label{lastpage}

\end{document}